\documentclass[fleqn,usenatbib]{mnras}
\usepackage[utf8]{inputenc}
\usepackage[T1]{fontenc}
\usepackage{graphicx}
\usepackage{amsmath}
\usepackage{amssymb}

\newcommand{\lambdabar}{{\mkern0.75mu\mathchar '26\mkern -9.75mu\lambda}}

\title[Pulsar death line revisited -- I]
{Pulsar death line revisited -- I. Almost vacuum gap}
\author[V. S. Beskin, P. E. Litvinov]
{V. S. Beskin$^{1,2}$\thanks{E-mail:
beskin@lpi.ru} and P. E. Litvinov$^{2}$ \\
$^{1}$P.N.Lebedev Physical Institute, Leninsky prosp., 53, Moscow, 119991, Russia\\
$^{2}$Moscow Institute of Physics and Technology, Dolgoprudny,
Moscow region, Institutsky per. 9, 141700, Russia}

\begin{document}

\date{Accepted 2021 December 3. Received 2021 October 23; in original form 2021 June 18}

%\pagerange{\pageref{firstpage}--\pageref{lastpage}} \pubyear{2021}

\maketitle

\label{firstpage}

\begin{abstract}
In this paper, which is the first in a series of papers devoted to a detailed 
analysis of ''the death line'' of radio pulsars, we consider a possibility of 
producing secondary particles at a sufficiently long pulsar period $P$. To this 
end, we reconsidered the potential drop necessary for secondary plasma generation 
in the inner gap over magnetic polar caps. Our research made it possible to refine 
the conditions for generating secondary plasma, such as the multiplicity of the 
production of secondary particles and their spatial distribution. Our 
research also made it possible to further quantitatively analyse the dependence 
of the possibility of secondary plasma generation on all parameters, including 
the inclination angle of the magnetic axis to the rotation axis, the polar cap 
size and the magnetic field geometry.
\end{abstract}

\begin{keywords}
stars: neutron – pulsars: general.
\end{keywords}

\section{Introduction}
\label{Sect1}

According to our modern understanding of the phenomenon of radio pulsars,  
their radio emission is associated  with secondary electron-positron plasma 
generated in the polar regions of a neutron star~\citep{sturrock, RS, Arons1982, 
L&GS, L&K}. It is therefore not surprising that the cessation condition of the 
generation of secondary particles is associated with the so-called ''death line'' 
on the $P$--${\dot P}$ (or $P$--$B_{0}$) diagram, where $P$ is the pulsar period,
and $B_{0}$ is the magnetic field at the magnetic pole. 

Detailed works devoted to the generation of secondary plasma have been underway 
since the beginning of the eighties~\citep{dauharding82, GI85, AE2002, Denis, 
Tim2010, ML2010, TimArons2013, PhSC15, TimHar2015, CPhS16}. 
Nevertheless, up to now, a large number of different options are discussed in
the literature~\citep{RS, blandford76, Arons1982, MelU}, 
leading to very different conditions which set the ''death line'' of radio 
pulsars~\citep{CR93, ZhHM2000, Hib&A, Faucher, KonarDeka}. 

In this and the following article, we set ourselves the task of reconsidering all 
basic approximations usually used in constructing the secondary 
plasma production model, but which may work poorly near ''the death line''. In 
particular, we assume that due to irregularity of the secondary plasma production, 
almost the entire region of open field lines can be 
considered in a vacuum approximation: $\rho_{\rm e} = 0$. In other words, 
our task is to clarify the position of the ''death valley''~\citep{CR93} using 
modern models of the magnetosphere structure and the deceleration of a neutron star; 
the latter is necessary to quantify the period derivative $ {\dot P} $ in 
terms of $P$, $B_{0}$, and the inclination angle $\chi$. The effects of general 
relativity will also be taken into account.

Please note that in this work, the ''classical'' mechanism of the particle 
production is considered due to the single-photon conversion of the $\gamma$-quantum 
into an electron-positron pair in a superstrong magnetic field. As is
well-known~\citep{sturrock, RS}, this process includes primary particle 
acceleration by a longitudinal electric field, emission of $\gamma$-quanta 
due to curvature radiation, production of secondary electron-positron pairs, 
and, finally, secondary particle acceleration in the opposite direction, 
which also leads to the creation of secondary particles. In other words, we do not 
consider particle production due to Inverse Compton Scattering, which, as is well 
known~\citep{blandford76, ZhHM2000}, can also be a source of hard $\gamma$-quanta. 
As an excuse, we note that we will first of all be interested in old pulsars, in
which the surface temperature may not be high enough to form a sufficient number
of X-ray photons. 

For the same reason, we do not take into consideration synchrotron 
photons emitted by secondary pairs. The point is that, as is well known 
(see, e.g.~\citealt{GI85, Denis} and Sect.~\ref{Sect3_1} below), the energy 
of synchrotron photons emitted by secondary particles is approximately 
$15$ --$20$ times less than the energy of curvature photons emitted by 
primary particles. Therefore, near the threshold of particle production, 
when the free path length of curvature photons become close to the radius 
of the star $R$, the pulsar magnetosphere appears to be transparent for 
synchrotron photons.

Let us recall that the cessation condition for the pair creation determining 
''the death line'' was first evaluated by~\citet{RS} from the equality 
of the height of the 1D vacuum gap 
\begin{equation}
     H_{\rm RS} \sim 
     \left(\frac{\hbar}{m_{\rm e}c}\right)^{2/7}
     \left(\frac{B_{0}}{B_{\rm cr}}\right)^{-4/7}
     R_{\rm L}^{3/7} R_{\rm c}^{2/7}
\label{HRS}    
\end{equation}
and the polar cap radius
\begin{equation}
    R_{\rm cap} \approx \left(\frac{\Omega R}{c}\right)^{1/2} R. 
\label{RRS}    
\end{equation}
Here $B_{\rm cr} = m_{\rm e}^2c^3/e\hbar = 4.4 \times 10^{13}$ G is
the Schwinger magnetic field, $R_{\rm L} = c/\Omega$ is the radius
of the light cylinder ($\Omega = 2\pi/P$ is the star angular velocity) 
and $R_{\rm c}$ is the curvature of the magnetic field lines near
the magnetic pole. For the magneto-dipole energy losses
\begin{equation}
    W_{\rm tot} \sim \frac{B_{0}^{2}\Omega^{4}R^{6}}{c^3}
\label{Wtot0}    
\end{equation}
and the dipole magnetic field stricture, when
\begin{equation}
   R_{\rm c} = \frac{4}{3} \, \frac{r}{\theta_{m}},
\label{Rc}    
\end{equation}
($r$ and $\theta_{m}$ are the polar coordinates relative to the magnetic axis, 
$r$ is the distance from the star center), one can obtain for ''the death line''~\citep{RS}
\begin{equation}
  {\dot P}_{-15} = \beta_{\rm d} P^{11/4},
\label{dl1}  
\end{equation}
where ${\dot P}_{-15} = 10^{15}{\dot P}$ and $\beta_{\rm d} \sim 1$.

It is clear that in the mid-70s such accuracy was quite acceptable,
especially since expression (\ref{dl1}) was really limited from below
most of the pulsars in the $P$--${\dot P}$ diagram. 
However, as was already shown by~\citet{CR93}, the standard RS model
(dipole magnetic field, etc.) gives a very large period derivative ${\dot P}$
for the observed ‘’death line’’. That is why the idea was put forward
to consider a more complex structure of the magnetic field
resulting in a decrease of the period derivative (and leading to the
appearance of a ''death valley''). However, as was already emphasized, the
observed period derivative ${\dot P}$ can be affected by many other reasons
(masses and radii of neutron stars, the size of the polar cap, the effects 
of general relativity), which were not considered in the qualitative
analysis carried out by~\citet{CR93}. In particular, within the framework 
of this model, the simplest magnetodipole formula was used to determine 
the period derivative ${\dot P}$, which, as it is now clear, does not 
correspond to reality. Therefore, one of the main tasks of our consideration 
is the question of what parameters can lead to a decrease in the observed
deceleration rate ${\dot P}$.

To clarify this issue, we consider an almost vacuum gap model of the polar 
region. Remember that all modern models of particle generation (including 
recent PIC simulations) implicitly assume free ejection of particles 
from the neutron star surface. Consequently, one could expect that the 
potential drop would be close to that predicted in the~\citet{Arons1982} 
model, i.e. much smaller than in the vacuum gap Ruderman-Sutherlend model. 
However, as was shown in recent works performed in the PIC 
simulation~\citep{Tim2010, PhTS20}, due to strong nonstationarity, vacuum
regions appear from time to time, with the potential drop being close to 
the vacuum gap model.
Particularly, such an assumption is natural for the pulsars located 
near ''the death line''. In this case, the beginning of the cascade (and, 
hence, the filling of this region with a secondary electron-positron 
plasma) can be initiated by the cosmic gamma background, which, as is 
known, leads to $10^5$--$10^8$ primary particles per second in the 
polar cap region~\citep{SRad82}.

On the other hand, for the pulsars in the vicinity of the 
''death line'', the free path length $l_{\gamma}$ of  $\gamma$-quanta leading 
to the production of secondary particles can be of the order of star radius $R$ 
(the scale of the diminishing of the dipole magnetic field), i.e. much 
larger than the transverse size of the polar cap $R_{\rm cap} \sim 0.01 R$. 
Therefore, we need to determine 3D potential resulting in the acceleration 
of primary particles.

Let us note straight away that below we consider only a dipole magnetic field despite a 
large number of works which indicated that it is impossible to explain ''the death
line'' in a dipole magnetic field~\citep{Arons93, AKh2002, BTs2010, IEP2016, Bilous}. 
In other words, one of our tasks is to verify the possibility of explaining 
the position of ''the death line'' by other factors which are usually not considered 
when analysing the processes of secondary plasma production. Among such possible 
factors, one can indicate a decrease in the magnetic field with distance from the 
star, the possibility of producing secondary pairs by $\gamma$-quanta whose energy
is much larger than the typically used characteristic energy of the maximum 
of the spectrum, as well as the fact that secondary plasma is generated on 
field lines located closer to the magnetic axis than gamma-ray emitting particles.
All these effects become significant near ''the death line'' when the free 
pass of $\gamma$-quanta becomes comparable to the radius of a neutron star.
The role of the non-dipole magnetic field and all the other physical
parameters which can affect the position of ''the death line'', will be discussed 
in detail in Paper II.

As for Paper I, which is the first in a series of papers devoted to 
a detailed analysis of ''the death line'' of radio pulsars, it is devoted to 
the possibility of producing secondary particles at sufficiently 
large periods $P$. In Sect.~\ref{Sect2}, we construct an exact three-dimensional 
solution for a longitudinal electric field $E_{\parallel}$ in the polar regions 
of a neutron star in the case when plasma is absent in the entire region 
of open field lines. We show that in real dipole geometry for non-zero 
inclination angles $\chi$, the longitudinal electric field decreases much slower 
than previously assumed. In addition, the corrections related to the effects 
of general relativity are determined. Further in Sect.~\ref{Sect3}, we show that 
close to ''the death line'' a major role in particle creation begin to play 
those $\gamma$-quanta whose energy is several times greater than the commonly
used characteristic energy of curvature radiation. Finally, in Sect.~\ref{Sect4},
the generation of secondary electron-positron pairs near ''the death line'' is 
considered when the second-generation particles produced by the conversion of 
synchrotron photons can be neglected. Sect.~\ref{Sect5} is devoted to the 
conclusion and discussion.

\section{Almost vacuum gap}
\label{Sect2}

\subsection{Potential drop}

To begin with, we revise the electric potential $\psi$ over the polar cap 
assuming that near ''the death line'', almost all plasma goes away so that one 
can put $\rho_{\rm e} = 0$ everywhere within the open magnetic field line region.
As in the rotation reference frame, the first Maxwell equation has a form~\citep{RS}
\begin{equation}
  \nabla \, {\bf E} = 4 \pi (\rho_{\rm e}-\rho_{\rm GJ}),
\end{equation}
where
\begin{equation}
\rho_{\rm GJ} = -\frac{{\bf \Omega B}}{2 \pi c}
\end{equation}
is the~\citet{GJ} charge density, the electric potential $\psi$ must 
satisfy the relation
\begin{equation}
    \nabla^2 \psi = - 2\frac{{\bf \Omega B}}{c} 
\label{1}    
\end{equation}
with the boundary conditions
\begin{eqnarray}
   \psi({\rm star \, surface}) = 0,  
\label{bound0} \\
   \psi({\rm separatrix}) = 0.
\label{bound1}
\end{eqnarray}
As is well-known (see, e.g.~\citealt{MHD} for more detail), the first boundary 
condition is related to the assumption of the high conductivity of a neutron 
star surface. The second condition on the separatrix separating the regions 
of open and closed field lines follows from the assumption that in the closed 
field line region, there is enough plasma to screen the longitudinal electric 
field. Note also that, despite the fact that the shape of the polar cap is not
circular at nonzero inclination angles, we restrict ourselves to considering
only the case of a circular polar cap. 

In real dipole geometry, when ${\bf B} = (3({\bf nm}){\bf n} - {\bf m})/r^3$, 
equation (\ref{1}) looks like
\begin{eqnarray}
\frac{1}{r^2} \frac{\partial}{\partial r} 
\left(r^2\frac{\partial\psi}{\partial r}\right) 
+ \frac{1}{r^2\sin\theta}\frac{\partial}{\partial \theta}
\left( \sin\theta\frac{\partial \psi}{\partial \theta}\right) 
+ \frac{1}{r^2\sin^2\theta}\frac{\partial^2 \psi}{\partial \varphi^2} 
\nonumber \\
= - 2\frac{\Omega B_{0}}{c} \, \frac{R^3}{r^3}\left(\cos\theta\cos\chi 
+ \frac{3}{2}\sin\theta\sin\varphi\sin\chi\right),
\label{Eq2}
\end{eqnarray}
where $\chi$ is the inclination angle. Then for the circular shape of the polar cap,
the boundary conditions (\ref{bound0})-- (\ref{bound1}) have a form
\begin{eqnarray}
    \psi(r=R, \theta, \varphi) & = & 0,
\label{gr21} \\
    \psi(r, \theta = \theta_{0}(r), \varphi) & = & 0,
\label{gr20}
\end{eqnarray}
where for small angles $\theta$
\begin{equation}
    \theta_{0}(r) = \left(\frac{rR_{0}^2}{R^3}\right)^{1/2}.
\end{equation}
Here
\begin{equation}
    R_{0} = f_{*}^{1/2} \, \left(\frac{\Omega R}{c}\right)^{1/2} R
\label{fstar}    
\end{equation}
is the polar cap radius, and we introduce standard dimensionless area $f \sim 1$.

For zero longitudinal electric currents, the dimensionless area of a polar cap $f_{*}$ 
changes from $f_{*} = 1.59$ for $\chi = 0^{\circ}$ to $f_{*} = 1.96$  for 
$\chi = 90^{\circ}$~\citep{michel73, bgi83}. Modern numerical simulations show that 
$f_{*}$ changes from $f_{*} = 1.46$ for $\chi = 0^{\circ}$ to $f_{*} = 1.75$  for 
$\chi = 90^{\circ}$~\citep{Gralla}. The convenience of introducing a dimensionless 
area $f$ is due to the fact that in a dipole field for small angles,
\begin{equation}
    \theta = f^{1/2} \, \left(\frac{\Omega r}{c}\right)^{1/2},
    \label{ftheta}
\end{equation}    
so that $f$ is constant along the dipole magnetic field line. 

As a result, the solution of equation (\ref{Eq2}) looks like
\begin{eqnarray}
&&\psi = \frac{1}{2} \, \frac{\Omega B_{0}R_{0}^2}{c}\cos\chi \times
\nonumber \\ 
&& \left[1 - \frac{\theta^2}{\theta_{0}^2(r)}  - \sum_{i} c_{i}^{(0)} \left(\frac{r}{R}\right)^{-\lambda_{i}^{(0)}/\theta_{0}'}J_{0}(\lambda_{i}^{(0)} \theta/\theta_{0}')\right]  
    \nonumber \\
&&    + \frac{3}{8} \, \frac{\Omega B_{0}R_{0}^2}{c}\, %\frac{R_{0}}{R}
\sin\varphi \sin\chi \times 
\label{7bis}\\
&& \left[\theta - \frac{\theta^3}{\theta_{0}^2(r)} - \theta_{0}' \sum_{i} c_{i}^{(1)} \left(\frac{r}{R}\right)^{-\lambda_{i}^{(1)}/\theta_{0}'}J_{1}(\lambda_{i}^{(1)} \theta/\theta_{0}')\right].
\nonumber 
\end{eqnarray}
Here  
\begin{equation}
    \sum c_{i}^{(0)}J_{0}(\lambda_{i}^{(0)} x) = 1 - x^2   
\label{sum1}
\end{equation}
and $\lambda_{i}$ are the zeros of the Bessel function $J_{0}(x)$: 
\begin{equation}
\lambda_{1}^{(0)} = 2.4, \, \lambda_{2}^{(0)} = 5.5, \, \lambda_{3}^{(0)} = 8.65, \, \lambda_{4}^{(0)} = 11.8, \dots
\end{equation}
Accordingly, $\lambda_{i}^{(1)}$ are the zeros of the Bessel function $J_{1}(x)$
\begin{equation}
\lambda_{1}^{(1)} = 3.8, \, \,  \lambda_{2}^{(1)} = 7.0, \, \, \lambda_{3}^{(1)} = 10.2, \, \, \lambda_{4}^{(1)} = 13.3, \dots
\end{equation}
and
\begin{equation}
    \sum c_{i}^{(1)}J_{1}(\lambda_{i}^{(1)} x) = x - x^3.   
\label{sum2}
\end{equation}
Since all the terms under the sum signs in (\ref{7bis}) rapidly
decrease with increasing $r$, we put here $\theta_{0}' = \theta_{0}(R) =$ const.

It is instructive to stress that for the symmetric ($\varphi$-independent) part, one can obtain
\begin{eqnarray}
    \sum c_{i}^{(0)} & = & 1, 
    \label{c00} \\
    \sum c_{i}^{(0)}(\lambda_{i}^{(0)})^2 J_{0}(\lambda_{i} x) & = & 4,
    \label{c01}
\end{eqnarray}
the last condition resulting from (\ref{1}). In particular, for $x = 0$,
we have
  \begin{eqnarray}
    \sum c_{i}^{(0)}(\lambda_{i}^{(0)})^2 & = & 4.
    \label{c01bis}
\end{eqnarray}  
  
Unfortunately, the second series (\ref{c01}) converges very slowly (see Appendix~\ref{A1}).
For this reason, in what follows, we restrict ourselves to only the first four
terms $c_{i}^{(0)}$ for which the coefficients $c_{1}^{(0)}$  and $c_{2}^{(0)}$ 
coincide with the precisely calculated values, and the two remaining ones are 
selected to satisfy relations (\ref{c00}) and (\ref{c01bis}). As a result, we obtain 
\begin{equation}
c_{1}^{(0)} = 1.09, \, c_{2}^{(0)} = -0.11, \, c_{3}^{(0)} = 0.028, \,
c_{4}^{(0)} = -0.0075,
\label{ccc}
\end{equation}
so that $\sum c_{i}^{(0)} = 1.00$ and $\sum c_{i}^{(0)}(\lambda_{i}^{(0)})^2 = 4.03$.
As shown in Appendix~\ref{A1}, in some respects, these four terms (\ref{ccc}) 
approximate the constant (\ref{c01}) even better than twenty exact coefficients 
$c_{i}^{(0)}$; in particular, the sum of the first twenty terms of the series 
(\ref{c01bis}) results in $3.37$ instead of $4$.

Accordingly, for anti-symmetric ($\varphi$-dependent) part, equation (\ref{1}) gives
\begin{equation}
    \sum c_{i}^{(1)}(\lambda_{i}^{(1)})^3 J_{1}(\lambda_{i} x) = 8x.   
\label{c11}
\end{equation}
    Together with (\ref{sum2}) %due to $J_{1}(x) \rightarrow x/2$ for 
    in the limit $x \rightarrow 0$ it gives 
\begin{eqnarray}
    \sum c_{i}^{(1)}\lambda_{i}^{(1)} & = & 2,
\label{c11bis}    \\
    \sum c_{i}^{(1)}(\lambda_{i}^{(1)})^3 & = & 16.   
\label{c11bis'}
\end{eqnarray}    
Restricting ourselves to the four terms of the series, we again 
determine the first two coefficients $c_{1}^{(1)}$ and $c_{2}^{(1)}$  
equal to the exactly calculated values (see Appendix~\ref{A1}) and the
two remaining ones due to the relations (\ref{c11bis}) and (\ref{c11bis'}). 
This approximation gives
\begin{equation}
c_{1}^{(1)} = 0.70, \, c_{2}^{(1)} = -0.14, \, c_{3}^{(1)} = 0.044, \, c_{4}^{(1)} = -0.0089, 
\label{cc}
\end{equation}
so that $\sum c_{i}^{(1)}\lambda_{i}^{(1)} = 2.00$ and 
$\sum c_{i}^{(1)}(\lambda_{i}^{(1)})^3 = 16.03$.
As shown in Appendix~\ref{A1}, the four terms (\ref{cc}) approximate the 
linear function (\ref{c11}) even better than the twenty exact coefficients 
$c_{i}^{(1)}$ not to say that the sum of the first twenty terms of the 
series (\ref{c11bis'}) results in $84$ instead of $16$.

As a result, we obtain for the potential $\psi(l)$ as a function of the distance
$l$ along the magnetic field line $f = $ const
\begin{eqnarray}
&&\psi(l)  =  \frac{1}{2} \, \frac{\Omega B_{0}R_{0}^2}{c}\cos\chi \times 
\nonumber \\ 
&&\left[1 - \frac{f}{f_{*}}  - \sum_{i} c_{i}^{(0)} \left(\frac{l}{R}\right)^{-\lambda_{i}^{(0)}/\theta_{0}}J_{0}(\lambda_{i}^{(0)} \sqrt{f/f_{*}})\right]  
    \nonumber \\
&&    + \frac{3}{8} \, \frac{\Omega B_{0}R_{0}^3}{cR} %\, \frac{R_{0}}{R}
\sin\varphi \sin\chi \left[\left(\frac{f}{f_{*}}\right)^{1/2} \left(1 - \frac{f}{f_{*}}\right)\left(\frac{l}{R}\right)^{1/2} \right.
\nonumber \\
&&  \left. -\sum_{i} c_{i}^{(1)} \left(\frac{l}{R}\right)^{-\lambda_{i}^{(1)}/\theta_{0}}J_{1}(\lambda_{i}^{(1)} \sqrt{f/f_{*}})\right].
\label{psipsi}
\end{eqnarray}
Here we can put $\theta_{0} = R_{0}/R = $ const.
Accordingly, the longitudinal electric field $E_{\parallel} = - \partial \psi/\partial l$
looks like
\begin{eqnarray}
&&E_{\parallel} = -\frac{1}{2} \, \frac{\Omega B_{0}R_{0}}{c}\cos\chi \times
\nonumber \\
&&\sum_{i} c_{i}^{(0)} \lambda_{i}^{(0)}\left(\frac{r}{R}\right)^{-\lambda_{i}^{(0)}/\theta_{0}-1}
J_{0}(\lambda_{i}^{(0)} \theta/\theta_{0})  
\label{Epar1}   \\
&&    - \frac{1}{4} \, \frac{\Omega B_{0}R_{0}}{c}\, \frac{R_{0}}{R}
\sin\varphi \sin\chi \times
\nonumber \\
&&\sum_{i} c_{i}^{(1)} \lambda_{i}^{(1)} \left(\frac{r}{R}\right)^{-\lambda_{i}^{(1)}/\theta_{0}-1}
J_{1}(\lambda_{i}^{(1)} \theta/\theta_{0})
\nonumber \\
&&-\frac{3}{16} \, \left(\frac{f}{f_{\ast}}\right)^{1/2} 
    \left(1 - \frac{f}{f_{\ast}}\right) \, \frac{\Omega B_{0}R_{0}^3}{c R^2} 
    \left(\frac{l}{R}\right)^{-1/2}  \sin\varphi \sin\chi.
\nonumber    
\end{eqnarray}

Here the following points should be stressed.
\begin{enumerate}
    \item 
The expression
\begin{eqnarray}
 \psi(r, \theta, \varphi)  = && \frac{1}{2} \, 
\frac{\Omega B_{0}R_{0}^2}{c}\left[1-\frac{\theta^2}{\theta^2_{0}(r)}\right]\cos\chi 
\label{psis} \\
&& + \, \frac{3}{8} \, \frac{\Omega B_{0}R_{0}^2}{c}\,  \left[\theta-\frac{\theta^3}{\theta^2_{0}(r)}\right]\sin\varphi \sin\chi
\nonumber
\end{eqnarray}
in (\ref{7bis}) is indeed the exact asymptotic solution at large distances
from the star surface $(r - R) \gg R_{0}$ in the dipole magnetic field
(certainly, in the limit $\theta \ll 1$). 
    \item 
At large distances $(r - R) \gg R_{0}$, the magnetic field lines $\theta(r) \propto r^{1/2}$ 
become equi-potential only for the symmetric component of the potential $\psi$. For the
anti-symmetric ($\varphi$-dependent) part realizing for any oblique rotator with
$\chi \neq 0^{\circ}$, the longitudinal electric field decreases on the scale of the star 
radius $R$, not the polar cap radius $R_{0}$. We emphasize that this effect does not 
exist for a model of the conical geometry of open magnetic field lines when
$\theta_{0} =$ const. 
    \item
As a result, one can obtain for the anti-symmetric component of the potential  
\begin{equation}
    \psi(l) = \frac{3}{8} \, \left(\frac{f}{f_{\ast}}\right)^{1/2} 
    \left(1 - \frac{f}{f_{\ast}}\right) \, \frac{\Omega B_{0}R_{0}^3}{c R} 
    \left(\frac{l}{R}\right)^{1/2}  \sin\varphi \sin\chi,
    \label{psiadd}
\end{equation}
where again $l$ denotes the coordinate along the magnetic field line,
and we use the dimensionless area $0 < f < f_{*}$. 
%For $\theta = f^{1/2} \, (\Omega l/c)^{1/2}$ (\ref{ftheta})
As we see, the additional potential drop within the light cylinder 
$\psi(l = R_{\rm L})$ is the same as the characteristic
vacuum potential within the polar cap.
    \item
Accordingly, the additional parallel electric field looks like 
\begin{equation}
    E_{\parallel}^{\rm add} = -\frac{3}{16} \, \left(\frac{f}{f_{\ast}}\right)^{1/2} 
    \left(1 - \frac{f}{f_{\ast}}\right) \, \frac{\Omega B_{0}R_{0}^3}{c R^2} 
    \left(\frac{l}{R}\right)^{-1/2}  \sin\varphi \sin\chi.
    \label{Eadd}
\end{equation}
    \item
Previously, no one paid attention to the existence of an additional longitudinal
field for the case of an oblique rotator when the dependence of the boundary of 
the region of open field lines on the distance to the star becomes significant. 
In particular, in the famous work of~\citet{MTs92}, a change of variables 
$\xi = \theta/\theta_{0}(r)$ was made when solving the equation (\ref{Eq2}), but in 
what follows, the dependence $\theta_{0}(r)$ on $r$ was not taken into account.
\end{enumerate}

\subsection{Particle acceleration}

The most important result obtained in the previous section is that, 
for the non-zero inclination angle $\chi$, there is the longitudinal 
electric field $E_{\parallel}$ (\ref{Eadd}) which decreases not on the 
polar cap scale $R_{0}$ but on the scale of the radius $r$ up to the
light cylinder. As a result, it produces a significant effect on 
the motion of particles, which, as is well-known, is described by the 
equation
\begin{equation}
\frac{{\rm d}{\cal E}_{\rm e}}{{\rm d}l} = eE_{\parallel}
-\frac{2}{3} \frac{e^2}{R_{\rm c}^2} \left(\frac{{\cal E}_{\rm e}}{m_{\rm e}c^2}\right)^4.
\label{motion}
\end{equation}
Here $R_{\rm c}$ is the curvature radius of the magnetic field line.
For $\theta \ll 1$ (i.e. for $l \ll R_{\rm L}$), one can put
\begin{equation}
R_{\rm c} \approx \frac{4}{3} f^{-1/2}R_{\rm L}^{1/2} l^{1/2}.
\end{equation}

\begin{figure}
%	\begin{minipage}{0.45\linewidth}
		\center{\includegraphics[width=1\linewidth]{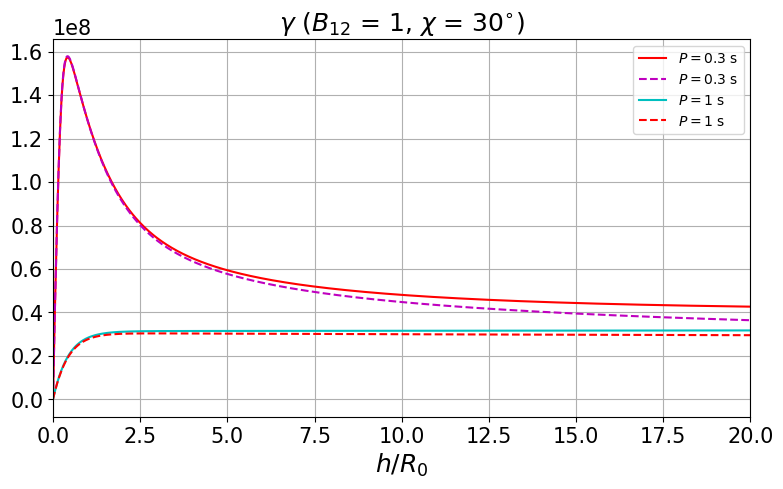}  }
%	\end{minipage}
	\hfill
%	\begin{minipage}{0.45\linewidth}
		\center{\includegraphics[width=1\linewidth]{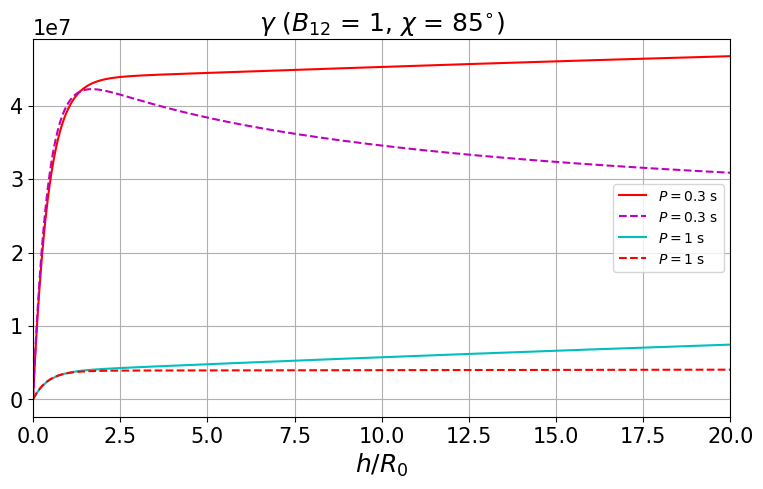}  }
%	\end{minipage}
	\caption{Lorentz-factor $\gamma = {\cal E}_{\rm e}/m_{\rm e}c^2$ of a 
	particle accelerated from the surface of a neutron star obtained by 
	solving the equation (\ref{motion}) for two values of the pulsar period 
	$P = 0.3$ s and  $P = 1$ s for small ($\chi = 30^{\circ}$, top) and large 
	($\chi = 85^{\circ}$, bottom) inclination angles. The dashed lines for 
	each of the two periods show the solutions in which the additional 
	electric field (\ref{Eadd}) is neglected.}
\label{FigD}	
\end{figure}

It should be immediately noted that due to the additional factor $R_{0}/R$, this 
component of the longitudinal electric field becomes significant near the 
surface of the neutron star for the inclination angles $\cos \chi \sim R_{0}/R$ 
only. Figure~\ref{FigD} shows the values of the Lorentz-factor 
$\gamma = {\cal E}_{\rm e}/m_{\rm e}c^2$ of a particle accelerated from the 
surface of a neutron star from the point $f = 0.7$ and $\varphi = 90^{\circ}$ 
obtained by solving the equation (\ref{motion}) for two values of the pulsar period 
$P = 0.3$ s (upper curves) and $P = 1$ s (lower curves) for small 
($\chi = 30^{\circ}$, top) and large ($\chi = 85^{\circ}$, bottom) inclination 
angles. The dashed lines show the solutions in which the additional electric field 
(\ref{Eadd}) is neglected. As we see, the role of the additional electric field 
becomes noticeable only for almost orthogonal rotators.

However, for us, it is much more important that for pulsars with relatively 
large periods $P \sim 1$ s, i.e. just near ''the death line'', the energy 
losses of a particle described by the second term in the equation (\ref{motion}) 
becomes negligible. Therefore, the lower curves in Figure~\ref{FigD} correspond 
to the electric potential $\psi(l)$. For this reason, in what follows, for 
such pulsars, we can simply put ${\cal E}_{\rm e}(l) = e \psi(l) - e\psi(l_{0})$,
where $l_{0}$  corresponds to the particle creation point. For shorter periods $P$, 
the particle energy does not reach the  maximum possible energy $e\psi$, and 
subsequently decreases with  increasing the distance from the neutron star surface.

\subsection{General relativistic correction}
\label{Sect2.3}

As is well-known, the effects of general relativity and, in particular, 
the frame-dragging  (Lense-Thirring) effect, under certain conditions, 
can play a significant role in the generation of secondary plasma near 
the polar caps of the neutron star~\citep{B90, MTs92, PhSC15, PhTS20}.
For this reason, below, we estimate all possible corrections which can 
affect the production of secondary particles. For simplicity, we restrict 
ourselves to only the first order in the small parameter $r_{\rm g} / R$, 
where $r_{\rm g} = 2 GM / c^2 $ is the black hole radius.

Starting from the time-independent Maxwell equation in the rotation reference
frame (see~\citealt{ThMc86} for more detail)
\begin{equation}
\nabla \times (\alpha {\bf E} + \bmath{\beta} \times {\bf B}  + \bmath{\beta}_{\rm R} \times {\bf B}) = 0,
\label{GR1}
\end{equation}
where $\alpha$ is the lapse function ($\alpha^2 \approx 1 - r_{\rm g}/R$), 
$\bmath{\beta}$ is Lense-Thirring vector and
$\bmath{\beta}_{\rm R} = \bmath{\Omega} \times {\bf r}/c$, we obtain
\begin{equation}
\alpha {\bf E} + \bmath{\beta} \times {\bf B}  + \bmath{\beta}_{\rm R} \times {\bf B} = - \nabla\psi.
\label{GR2}
\end{equation}
For $\rho_{\rm e} = 0$, this equation gives
\begin{equation}
\nabla \left(\frac{\nabla \psi}{\alpha}\right) = 4 \pi \rho_{\rm GJ},
\label{GR3}
\end{equation}
where now the Goldreich-Julian charge density looks like
\begin{equation}
\rho_{\rm GJ} = -\frac{1}{8 \pi^2} \nabla_{k}\left(\frac{\Omega - \omega}{\alpha c}\nabla^{k} \Psi\right). 
\label{GR4}
\end{equation}

As we see, the first relativistic correction \mbox{$(1 - \omega/\Omega)$} 
appears in the expression for $\rho_{\rm GJ}$, where the ratio $\omega/\Omega$ 
depends on the neutron star moment of inertia $I_{r} \sim MR^2$:
\begin{equation}
\frac{\omega}{\Omega} = \frac{I_{r}r_{\rm g}}{M r^3}.
\label{GR5}
\end{equation}
Thus this correction just corresponds to a small value $\sim r_{\rm g}/R$ under consideration. To determine the relativistic corrections, we chose the values $M = 1.4 \, M_{\odot}$ for neutron star mass, $R = 13$ km for that radius and $I_{r} = 150$ $M_{\odot}$km$^{2}$ for the moment of inertia~\citep{Greif20}. Notably, the characteristic scale of the changes in all the relativistic corrections are $R$, in contrast, the scale of change in $\psi$ is $R_{0} \ll R$. Therefore, one can consider all relativistic corrections as constants, i.e. we can put $r = R$ in (\ref{GR5}). The second relativistic correction appears in the expression for the magnetic field flux
\begin{equation}
\Psi = 2\pi |{\bf m}|\frac{\sin^2\theta_{m}}{r}\left(1 + \frac{3}{4}\frac{r_{\rm g}}{r}\right).
\label{GR6}
\end{equation}
As for small angles $\theta_{m}$, one can put $\sin\theta_{m} = r_{\perp}/r$, i.e. to write down
\begin{equation}
x^2 = \frac{\Psi}{2 \pi |{\bf m}|}
y^3\left(1 + \frac{3}{4}\frac{r_{\rm g}}{y}\right)^{-1},
\label{GR7}
\end{equation}
where $x = r_{\perp}$ and $y = r$,  
we obtain for the curvature radius $R_{\rm c} \approx 1/y''_{xx}$ the following correction $R_{\rm c, GR} = K_{\rm cur}R_{\rm c}$, where 
\begin{equation}
K_{\rm cur} = \left(1 - \frac{1}{2}\frac{r_{\rm g}}{R}\right).
\label{GR8}
\end{equation}
Next, for the polar cap radius $R_{0, GR} = K_{\rm cap}R_{0}$, we have
\begin{equation}
K_{\rm cap} = \left(1 - \frac{3}{8}\frac{r_{\rm g}}{R}\right).
\label{GR9}
\end{equation}
Finally, Eqn. (\ref{GR3}) looks now like
\begin{equation}
\frac{\alpha^2}{r_{\perp}}\frac{\partial}{\partial r_{\perp}}\left(r_{\perp}\frac{\partial \psi}{\partial r_{\perp}}\right) 
+ \frac{\partial^2 \psi}{\partial z^2} = 
- \frac{2 \Omega B_{0}}{c} \left(1 + \frac{3}{4} \frac{r_{\rm g}}{R}\right)\left(1 - \frac{\omega}{\Omega}\right).
\label{GR10}
\end{equation}
As a result, we obtain the general relativistic correction for the symmetric potential \mbox{$\psi_{GR}(r_{\perp}) =  K_{\psi}\psi(r_{\perp})$} at distances $h > R_{0}$ over the star surface as
\begin{equation}
 K_{\psi} = \left(1 - \frac{\omega}{\Omega}\right)\left(1 - \frac{r_{\rm g}}{R}\right)^{-1}.
\label{GR11}
\end{equation}

Thus, as expected, we can conclude that the effects of general relativity lead to corrections at the 
level of 10--20 per cent. Consequently, the corresponding corrections at first glance  do not go beyond 
the uncertainty in other quantities, such as, e.g. the radius and moment of inertia of a neutron star.
Nevertheless, below, we include general relativity corrections into consideration since, as will be shown,
these corrections lead to a significant change in the rate of production of secondary particles.

\section{Generation of curvature photons}
\label{Sect3}

\subsection{Effective photon energy}
\label{Sect3_1}

The next point, which can be important in the vicinity ''the death line'' 
is the question of the effective energy of the curvature \mbox{$\gamma$-quanta} 
leading to the production of the secondary pairs. Recall that in
most cases (see, e.g.~\citealt{TimHar2015}), it was assumed that 
all the photons emitted by the relativistic electron with the energy 
${\cal E}_{\rm e} = \gamma_{\rm e}m_{\rm e}c^2$ are radiated at 
the characteristic frequency
\begin{equation}
    \omega_{\rm c} = \frac{3}{2} \frac{c}{R_{\rm c}} \gamma_{\rm e}^3,
    \label{41}
\end{equation}
which is larger than the maximum of the energy spectrum 
$0.29 \, \omega_{\rm c}$~\citep{LL}. Below, we show that the very 
first pairs are produced by even more energetic photons, whose 
frequencies are several times higher than the characteristic 
frequency $\omega_{\rm c}$ (\ref{41}).

Indeed, the spectrum of the curvature radiation (i.e. the energy
radiated in the frequency domain ${\rm d}\omega$ at the distance
${\rm d}l$){\footnote{This expression can be easily obtained from 
a well-known synchrotron spectrum~\citep{LL} by replacing the Larmor 
radius $r_{\rm L} = m_{\rm e}c^2 \gamma/eB$ with the curvature radius  
of the magnetic field line $R_{\rm c}$.}}
\begin{equation}
    {\rm d}I = \frac{\sqrt{3}}{2 \pi} \frac{e^2}{c R_{\rm c}} 
    \gamma_{\rm e} F(\omega/\omega_{\rm c})  {\rm d} \omega \, {\rm d}l,
    \label{42}
\end{equation}
where
\begin{equation}
    F(\xi) = \xi\int_{\xi}^{\infty}K_{5/3}(x){\rm d}x,
    \label{43}
\end{equation}
$K_{5/3}$ is the Macdonald function, and $\xi = \omega/\omega_{\rm c}$, 
has a rather long tail. As a result, although a relativistic particle 
needs to travel a certain distance $l_{\rm rad}$ for the emission of 
high-energy photons with $\omega \gg \omega_{\rm c}$, the free path 
length $l_{\gamma}$ of a $\gamma$-quantum before the creation of a 
secondary electron-positron pair will be much shorter than for the
photons radiated near the maximum. 

Since in what follows, we are only interested in the photons with frequencies 
$\omega \gg \omega_{\rm c}$, one can use the following asymptotic behaviour 
of the Macdonald's function~\citep{AS}
\begin{equation}
    K_{5/3}(x) \approx \sqrt{\frac{\pi}{2 x}}\, e^{-x}\left(1
    + \frac{91}{72}\frac{1}{x} + \dots \right),
    \label{44}
\end{equation}
i.e. consider only the two first terms of the expansion in terms of $1/x$. 
Accordingly, with the same accuracy
\begin{equation}
    F(x) \approx \sqrt{\frac{\pi x}{2}}\, e^{-x}\left(1
    + \frac{55}{72}\frac{1}{x} + \dots \right).
    \label{44'}
\end{equation}

Below, we need a general expression for an arbitrary dependence of the energy 
${\cal E}_{\rm e}(l) = \gamma_{\rm e}(l) m_{\rm e}c^2$  of the emitting 
particle on the distance $l$. Assuming that a photon with the frequency 
$\omega \gg \omega_{\rm c}$ can be emitted if the total energy
\begin{equation}
{\cal E}_{\omega} = \frac{\sqrt{3}}{2 \pi} \frac{e^2}{c R_{\rm c}} 
\int_{l_{0}}^{l_{0} + l_{\rm rad}} \,\int_{\omega}^{\infty} 
    \gamma_{\rm e}(l) F[\xi(l)]  \, {\rm d}  \omega \, {\rm d} l , 
\label{45'}
\end{equation}
radiated above this frequency is equal to photon energy: 
${\cal E}_{\omega} = \hbar \omega$. Here, 
$\xi(l) = (3/2) (c/R_{\rm c}) \gamma_{\rm e}^{3}(l)$, and $l_{0}$ 
is the radiation coordinate. For the constant particle energy 
$\gamma_{\rm e} =$ const, we have 
\begin{equation}
{\cal E}_{\omega} = \frac{\sqrt{3}}{2 \pi} \frac{e^2}{c R_{\rm c}} 
\omega_{\rm c} \, l_{\rm rad} \, \gamma_{\rm e} \,\int_{\xi}^{\infty} 
     F(\xi)  {\rm d} \xi . 
\label{45}
\end{equation}

Introducing finally a new variable
\begin{equation}
\xi^{\prime} = \frac{2}{3} \, \frac{\omega R_{\rm c}}{c}  \gamma_{\rm e}^{-3}(l),% \frac{\gamma_{\rm e}^{3}(l_0)}{\gamma_{\rm e}^{3}(l_0)},
\end{equation}
we obtain 
\begin{equation}
\xi = \frac{\sqrt{3}}{2 \pi} \frac{e^2}{\hbar c} 
\int_{0}^{l_{\rm rad}} \frac{{\rm d} l}{R_{\rm c}}\,
\gamma_{\rm e}(0)\frac{\gamma_{\rm e}^{4}(l)}{\gamma_{\rm e}^{4}(0)}\int_{\xi}^{\infty} 
     F(\xi^{\prime})  \, {\rm d}  \xi^{\prime},
\label{45''}
\end{equation}
where $\gamma(0) = \gamma(l_{0})$ and $\gamma(l_{\rm rad}) = \gamma(l_{0} + l_{\rm rad})$. 
Relation (\ref{45''}) determines implicitly the connection between $l_{\rm rad}$ and $\xi$.
In particular, for $\gamma_{\rm e} =$ const, we obtain for the radiation length $l_{\rm rad}$
\begin{equation}
l_{\rm rad} = \sqrt{\frac{8 \pi}{3}}  \, \frac{\hbar c}{e^2} \, R_{\rm c} \, \gamma_{\rm e}^{-1}
\sqrt{\xi} \left(1 - \frac{91}{72} \, \frac{1}{\xi} + \dots \right) e^{\xi}.
    \label{46}
\end{equation}

Further, the free path length $l_{\gamma}$ of a photon can be written as~\citep{sturrock, RS}
\begin{equation}
l_{\gamma} = \frac{8}{3\Lambda} \,R_{\rm c} 
\frac{B_{\rm cr}}{B} \frac{m_{\rm e}c^2}{\hbar \omega_{\rm c}} \, \frac{1}{\xi},
    \label{47}
\end{equation}
where $\Lambda = 15$--$20$ is 
the logarithmic factor: $\Lambda \approx \Lambda_{0} - 3\ln\Lambda_{0}$, where
\begin{equation}
\Lambda_{0} = \ln\left[
\frac{e^2}{\hbar c}\,\frac{\omega_B R_{\rm c}}{c}
\left(\frac{B_{\rm cr}}{B}\right)^2
\left(\frac{m_{\rm e}c^2}{{\cal E}_{\rm ph}}\right)^2\right] \sim 20.
   \label{48}
\end{equation}

\begin{table}
\caption{Tabulation of the inverse function $\xi({\cal K})$ (\ref{49}).}
\begin{tabular}{ccccccccc}
\hline
${\cal K}$ & 30 & 100 & 300 & $10^3$ & $3\cdot 10^3$ & $10^4$ & $3 \cdot 10^4$ &  $10^5$ \\
$\xi$ & 2.1 & 2.6 & 3.1 & 3.8 & 4.5 & 5.2 & 6.0  & 6.8 \\
\hline
\end{tabular}
\label{tab1}
\end{table}

Minimizing now the sum $l_{\rm rad} + l_{\gamma}$ by the value $\xi$, one can
obtain the energy of a photon $\hbar \omega = \xi \hbar \omega_{\rm c}$ producing 
the first secondary pair. In particular, for the constant particle energy, we have 
the following relation to determine the value $\xi$
\begin{equation}
\xi^{5/2} \, e^{\xi} \left(1 - \frac{55}{72} \, \frac{1}{\xi} + \dots \right) = {\cal K},
    \label{49}
\end{equation}
where 
\begin{equation}
{\cal K} = \frac{4\sqrt{2}}{3\sqrt{3 \pi} \Lambda} \, \frac{B_{\rm cr}}{B}  
 \, \frac{R_{\rm c}}{a_{\rm B}}  \, \gamma_{\rm e}^{-2} \approx
 40 \, R_{\rm c, 7} B_{12}^{-1}\gamma_{7}^{-2}. 
    \label{50}
\end{equation}
Here, $B_{12} = B/(10^{12}$ G), $R_{\rm c, 7} = R_{\rm c}/(10^{7}$ cm), 
$\gamma_{7} = \gamma_{\rm e}/10^{7}$, and 
$a_{\rm B} = \hbar^2/m_{\rm e}e^2 = 5.3 \times 10^{-9}$ cm is the 
Bohr radius. Accordingly, the total length $l_{\rm tot} = l_{\rm rad} + l_{\gamma}$
in this case looks like
\begin{equation}
l_{\rm tot} = \frac{8}{3\Lambda} \, \frac{B_{\rm cr}}{B} 
\frac{m_{\rm e}c^2}{\xi \hbar \omega_{\rm c}}\left(1 + \frac{1}{\xi}\right). 
\label{50'}
\end{equation}

As shown in Table~\ref{tab1}, even for the characteristic parameters, 
the most effective frequency $\omega = \xi \omega_{\rm c}$  is indeed 
higher than $\omega_{\rm c}$ (\ref{41}). 
As for the pulsars located within ''the death valley'', their effective frequency
can be sufficiently higher. In Table~\ref{tab0}, we show the parameters ${\cal K}$ 
and $\xi$ for some of these pulsars. Their parameters were taken
from the ATNF catalogue~\citep{ATNF}, and the appropriate Lorentz-factors correspond to 
the potential drop $\psi$ (\ref{psipsi}). As we see, for all these pulsars, the effect
under consideration can indeed play a significant role. A detailed analysis of all 
the pulsars located within ''the death valley'' will be provided in Paper II.

\begin{table}
\caption{Values ${\cal K}$ and $\xi$ for some pulsars located within ''the death valley''. 
The pulsar parameters are taken from the ATNF catalogue~\citep{ATNF}.}
\begin{tabular}{cccccc}
\hline
PSR & $P$ (s) & ${\dot P}_{-15}$ &  $B_{12}$ & ${\cal K}$  & $\xi$   \\
\hline
J0250$+$5854    &  23.5 &  27.1 &  25  &  $3.3 \times 10^4$  & 6.1 \\
J0418$+$5732    &  9.01  &  4.10  &  6.1 &  $3.2 \times 10^4$  & 6.1 \\
J1210$-$6550    &  4.24  &  0.43  & 1.3  &  $1.0 \times 10^5$  & 6.8 \\
J1333$-$4449    &  0.46  &  0.0005  &  0.016 & $3.3 \times 10^6 $ &  9.5 \\
J2144$-$3933    &  8.51  &  0.50  & 2.1  &  $6.6 \times 10^5$  &  8.2  \\
J2251$-$3711    &  12.1  &  13,1  &  12.5 & $1.4 \times 10^4$ & 5.5  \\
\hline
\end{tabular}
\label{tab0}
\end{table}

\subsection{Free pass}

When determining the value of $\xi$, the accuracy in determining the logarithmic 
factor $\Lambda$ is insignificant since equation (\ref{49}) actually gives
$\xi \sim \ln K$. On the other hand, near ''the death line'', more accurate determination
of the free path length $l_{\gamma}$ is necessary. This, in particular, is due to 
the fact that $l_{\gamma}$ turns out to be comparable with the radius of the star 
$R$, i.e. with the scale at which the magnetic field of the neutron star decreases 
significantly; this was not taken into account when deriving equation (47).
For this reason, to determine the path length of a photon, we will use the exact 
expression for the probability $w_{l}$ of the photon production at a length  
${\rm d}l$~\citep{BLP}
\begin{equation}
{\rm d}w_{l} = \frac{3 \sqrt{3}}{16 \sqrt{2}} \,
\frac{e^3 B\sin\theta_{\rm b}}{\hbar m_{\rm e}c^3}
\exp\left(-\frac{8}{3}\frac{B_{\rm cr}}{B(l)\sin\theta_{\rm b}(l)}
\frac{m_{\rm e}c^2}{{\cal E}_{\rm ph}}\right){\rm d}l,
\label{51}
\end{equation}
where again $B_{\rm cr} = m_{\rm e}^2c^3/e\hbar \approx 4.4 \times 10^{13}$  G is the 
critical magnetic field, and $\theta_{\rm b}$ is the angle between the magnetic 
field ${\bf B}$ and the wave vector ${\bf k}$. As a result, the free pass length $l_{\gamma}$
should be determined from the condition
\begin{equation}
\int_{0}^{l_{\gamma}}{\rm d}w_{l} = 1.
\label{52}
\end{equation}

In conclusion, we note one more convenience of using the  parameter
$\Lambda $. Indeed, since in a not very strong magnetic field 
$B_{0} \sim 10^{12}$ G, secondary particles at the time of birth 
move at the angle $\theta \sim l_{\gamma}/R_{\rm c}$ to the magnetic 
field lines, while their energy $\gamma_{\rm} m_{\rm e}c^2$ is 
to be  one half of the energy of the curvature photon 
$\hbar \, \omega_{\rm cur}$~\citep{VBAstro, DH83}, the ratio of 
the characteristic frequency of a synchrotron photon 
$\omega_{\rm syn} = (3/2) \theta \omega_{B}\gamma^2$ to the 
frequency of a curvature photon $\omega_{\rm cur}$ 
\begin{equation}
\frac{\omega_{\rm syn}}{\omega_{\rm cur}} = \frac{3}{8}\frac{B_{0}}{B_{\rm cr}}
\frac{\hbar \, \omega_{\rm cur}}{m_{\rm e}c^2}\frac{l_{\gamma}}{R_{\rm c}}
\label{52bis}
\end{equation}
turns out to be exactly $\Lambda^{-1}$. Just for this reason, as was already
mentioned above, free path length of synchrotron photons is to be $15$--$20$ 
times larger than free path length of the curvature photons. Therefore, 
near ''the death line'', the role of synchrotron photons should not be significant.

\section{Generation of secondary pairs}
\label{Sect4}

\subsection{Outflow}

In general, we follow the approach developed by~\citet{Hib&A}. The main difference 
is that we analyse the dependence of the number density of secondary particles $n_{\pm}$ 
on the distance $r_{\perp}$ from the magnetic axis rather than the energy spectrum.

Let us consider one primary particle moving along the magnetic field line intersecting the 
star surface at the distance $r_{0}$ from the axis. It produces ${\rm d}N$ curvature photons 
in the frequency domain $ {\rm d}\omega$ at the path ${\rm d}h$
\begin{equation}
 {\rm d}N =  \frac{\sqrt{3}}{2 \pi}\frac{e^2}{c R_{\rm c}(h)} 
    \frac{\gamma_{\rm e} F(\omega/\omega_{\rm c})}{\hbar \omega} 
%    \frac{{\rm d}\omega}{{\rm d}r_{\perp}} \, {\rm d} r_{\perp} 
  {\rm d}\omega  \, {\rm d}h.
\label{}
\end{equation}
On the other hand, frequency $\omega$ determines the free path 
$l_{\gamma} = l_{\gamma}(\omega)$ which, in turn,  determines 
the magnetic field line at which the secondary pair is created  
\begin{equation}
r_{\perp} = \left(1 - \frac{3}{8} \,\frac{l_{\gamma}^2}{R^2} \right)r_{0},
\label{xx}
\end{equation}
where again $r_{\perp}$ is the distance from the axis at the star surface, and  $h$ is the height of 
the $\gamma$-quanta emission.  Note that in a dipole magnetic field, this expansion does not contain 
the corrections $\propto hl_{\gamma}/R^2$ (and, certainly, it does not contain the term 
$\propto h^2/R^2$ as $r_{\perp} = r_{0}$ for $l_{\gamma} = 0$). 

\begin{table}
\caption{Tabulation of the function ${\cal L}(x_{0}, x_{\perp}, h)$ (\ref{calL}) for $x_{0} = 0.6$ and for different values $x_{\perp}$.}
\begin{tabular}{ccccccccc}
\hline
$h/R$ & 0.0 & 0.1 & 0.2 & 0.3 & 0.4 & 0.5 & 0.6 & 0.7 \\
\hline
0.599  & 1.0  & 1.4 & 1.9 & 2.6 & 3.4 & 4.3 & 5.4 & 6.7  \\ 
0.59  & 1.5 & 2.0 & 2.7  & 3.4 & 4.3 & 5.4 & 6.6  & 8.1 \\ 
0.58  & 1.9 & 2.5 & 3.3  & 4.1 & 5.1 & 6.3 & 7.7  & 9.3 \\ 
\hline
\end{tabular}
\label{tab2}
\end{table}

As we show below, the leading term in (\ref{xx}) is enough for our consideration. On the other hand, 
in what follows, to determine with the required accuracy the exponent in the pair creation probability 
$w_{l}(\theta_{\rm b})$ (\ref{51}), we use the exact expression for the angle $\theta_{\rm b}$ between
 the magnetic field {\bf B} and the wave vector ${\bf k}$. In a dipole magnetic field, for $\gamma$-quanta 
radiated tangentially at the height $h$,\ it looks like 
\begin{equation}
\theta_{\rm b} = \frac{3}{4} \, \frac{r_{0}l_{\gamma}}{R^2} f(h),
\label{tb}
\end{equation}
where the correction function $f(h)$ for $\theta_{\rm b} \ll 1$ is
\begin{equation}
f(h) = \left(1 + \frac{h}{R}\right)^{1/2}\left(1 + \frac{{\cal L}(h) l_{0}}{R} + \frac{h}{R}\right)^{-1}.
\label{fh}
\end{equation}
Here, we introduce by definition another correction function ${\cal L}(h)$ as
\begin{equation}
l_{\gamma}(l_{0}, h)  = {\cal L}(h) \,l_{0},
\label{calL}
\end{equation}
where 
\begin{equation}
l_{0}(\omega) = \frac{32}{9\Lambda} \, \frac{R^2}{r_{0}}
\frac{B_{\rm cr}}{B_{0}} \frac{m_{\rm e}c^2}{\hbar \omega}
\label{}
\end{equation}
is the $\gamma$-quantum free pass length in the case $l_{\gamma} \ll R$ with the starting point $h = 0$. The coefficient ${\cal L}(h)$ due to the strong nonlinearity of the problem for $h \sim l_{0} \sim R$ should be determined numerically by direct integration (\ref{52}) 
\begin{equation}
\int_{0}^{l_{\gamma}}w_{l}(h)\, {\rm d}l = 1
\label{}
\end{equation}
for probability $w_{l}(h)$ corresponding to the starting point $h$ at which the photon free pass is equal 
to $l_{\gamma}$ (see Table 2). Certainly, ${\cal L} \rightarrow 1$ for $h \rightarrow 0$ and 
$l_{\gamma} \rightarrow 0$ ($r_{\perp} \rightarrow r_{0}$). Finally, for the primary particle moving 
along the magnetic field line, we have 
\begin{equation}
R_{\rm c} = \frac{4}{3}\, \frac{R^2}{r_{0}} \left(1 + \frac{h}{R}\right)^{1/2}.
\label{}
\end{equation}

As a result, one can write down for the $r_{\perp}$ distribution of the secondary particles as
\begin{equation}
{\rm d}N  = \frac{\sqrt{3}}{2 \pi} \frac{e^2}{\hbar c} 
    \frac{\gamma_{\rm e} F(\omega/\omega_{\rm c})}{R_{\rm c} \omega} 
    \frac{{\rm d}\omega}{\,{\rm d}r_{\perp}}{\rm d}r_{\perp}
 \, {\rm d}h.
\label{xxx}
\end{equation}
To determine the derivative ${\rm d}\omega/{\rm d}r_{\perp}$, one can rewrite the relation (\ref{xx}) as
\begin{equation}
\frac{l_{\gamma}(\omega)}{R} = \frac{2\sqrt{2}}{\sqrt{3}} \, \frac{(r_{0}-r_{\perp})^{1/2}}{r_{0}^{1/2}}.
\label{lgxx}
\end{equation}
It finally gives 
\begin{equation}
\frac{1}{\omega} \frac{{\rm d}\omega}{{\rm d}r_{\perp}} = \frac{1}{2(r_{0}-r_{\perp})}\left(1 -  \frac{\omega}{\cal L}\frac{{\rm d}{\cal L}}{{\rm d}\omega}\right)^{-1}.
\label{}
\end{equation}
Within approximation (\ref{xx}) we consider, the value of ${\cal L}$ does not depend on $\omega$ 
(both free path length $l_{0}$ and $l_{\gamma}$ are mainly determined by the exponent, which both 
depend on $\omega$ as $\omega^{-1}$); therefore, below, we do not take into account the logarithmic derivative 
$\omega/{\cal L} ({\rm d}{\cal L}/{\rm d}\omega)$.

As a result, we obtain for the linear distribution of secondary particle $n_{\pm} \, {\rm d}x_{\perp}$ creating 
by one primary particle moving along the magnetic field line with a foot point distance from the axis $r_{0}$
\begin{equation}
n_{\pm}(r_{\perp}) = \frac{3\sqrt{3}}{16 \pi} \frac{e^2}{\hbar c}  \frac{R_{0}}{R}  \frac{x_{0}}{(x_{0}-x_{\perp})}
\int_{0}^{H}\frac{{\rm d}h}{R}\gamma_{\rm e}(x_{0},h) F(\xi),
\label{nescd1}
\end{equation}
where now
\begin{equation}
\xi = \frac{64\sqrt{2}}{27\sqrt{3}\Lambda}
\frac{B_{\rm cr}}{B_{0}} \frac{R^3}{\lambdabar R_{0}^2}\frac{\left(1 + {h}/{R} \right)}{\gamma_{\rm e}^3(x_{0},h)}\frac{{\cal L}(x_{0}, x_{\perp}, h)}{x_{0}\sqrt{x_{0}}\sqrt{x_{0}-x_{\perp}}} .
\label{xi1}
\end{equation}
Here, we introduce by definition two dimensionless parameters 
\begin{equation}
x_{0} = \frac{r_{0}}{R_{0}}; \qquad x_{\perp} = \frac{r_{\perp}}{R_{0}}.
\label{nesc}
\end{equation}
As for the upper integration limit $H$, it can be set equal to infinity, since, as shown in Table~\ref{tab2}, 
the parameter ${\cal L}$ introduced above increases rapidly with increasing $h$. Therefore, already at 
$h \sim H$, the integrand becomes exponentially small due to the large value of the argument $\xi$ (\ref{xi1}). 

\begin{figure}
		\center{\includegraphics[width=0.9\linewidth]{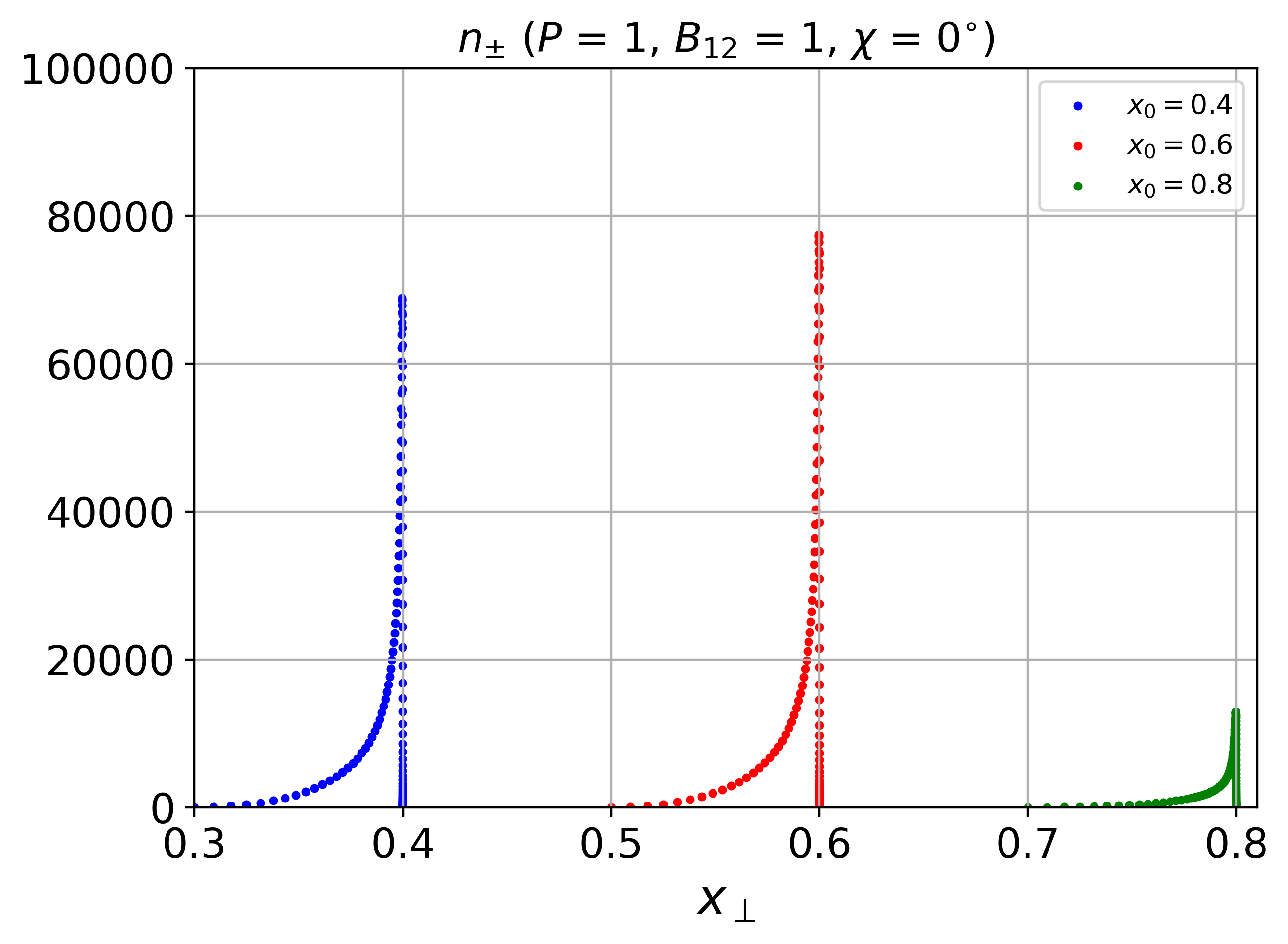}  }
	\caption{Secondary particle distribution $n_{\pm}(r_{\perp})$ (\ref{nescd1}) generated by a single primary particle with the starting points $x_{0} = 0.4$, $x_{0} = 0.6$ and $x_{0} = 0.8$ for $P = 1$ s, $B_{12} = 1$ and $\chi = 0^{\circ}$.}
\label{FigE1}	
\end{figure}

\begin{figure}
		\center{\includegraphics[width=0.9\linewidth]{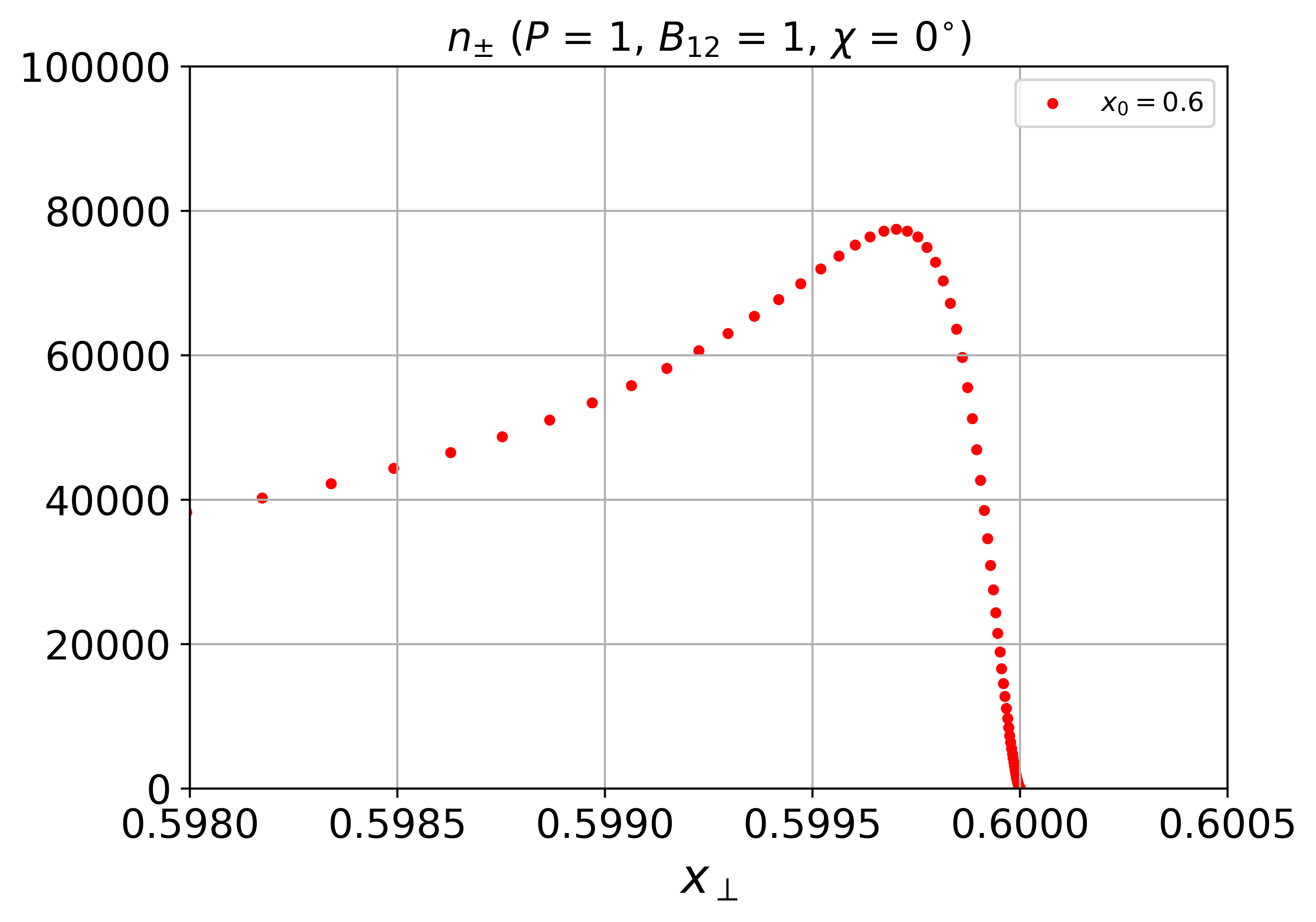}  }
	\caption{Secondary particle distribution $n_{\pm}(r_{\perp})$ (\ref{nescd1}) generated by a single primary particle with the starting points $x_{0} = 0.6$ near the threshold $x_{\perp} = x_{0}$.}
\label{FigE2}	
\end{figure}

In Figure~\ref{FigE1}, we show secondary particle distributions $n_{\pm}(r_{\perp})$ (\ref{nescd1}) generated 
by single primary particles with the starting points $x_{0} = 0.4$, $x_{0} = 0.6$ and $x_{0} = 0.8$ for 
\mbox{$P = 1$ s,} $B_{12} = 1$ and $\chi = 0^{\circ}$. The energy of the primary particles 
${\cal E}_{\rm e} = \gamma_{\rm b } m_{\rm e}c^2$ was determined from the vacuum potential 
$\psi$ (\ref{psipsi}). As we see, even though the maximum of the distribution $n_{\pm}(r_{\perp})$ 
lies near $r_ {0} $ (i.e. the corresponding secondary particles are born on practically the same 
field line as the primary particle), this distribution slowly decreases with increasing distance 
$r_{0} - r_{\perp}$. Consequently, with the parameters considered here, which are close to 
''the death line'', a significant part of the secondary particles will be produced at distances 
$h$, comparable to the radius of the star $R$. That is why we do not consider further the 
production of secondary plasma associated with the synchrotron radiation of secondary particles.

On the other hand, it is clear that $n_{\pm} = 0$ for $x_{\perp} = x_{0}$. As for the position of the maximum in the distributions $n_{\pm}$, they can be easily determined by setting argument (\ref{xi1}) as $\xi = 1$ so that $F(\xi)$ becomes exponentially small for $\xi > 1$. Moreover, using the potential $\psi$ (\ref{psipsi}) to determine $\gamma_{\rm b}$, one can write down the condition $\xi = 1$ as 
\begin{equation}
(x_{0} - x_{\perp}) = x_{0}^{-3} {\cal A},
\label{xi20}
\end{equation}
where %for $gamma_{\rm b}$ from (\ref{})
\begin{equation}
{\cal A} = \frac{2^7}{3^9 \, \Lambda^2} \, \frac{B_{\rm cr}^2}{B_{0}^2} \, 
\frac{R^6}{\lambdabar^{2}R_{0}^4} \gamma_{\rm b}^{-6} \approx 6 \times 10^{-5} P^{14} B_{12}^{-8}.
\label{xi21}
\end{equation}
Here, we took into account that ${\cal L} \approx 1$ for $x_{\perp} \rightarrow x_{0}$. Therefore, 
for $P = 1$ s and $B_{12} = 1$ (and for $x_{0} = 0.6$), we obtain $x_{0} - x_{\rm max} \approx 0.003$. 
As shown in Figure~\ref{FigE2}, this evaluation is in good agreement with our numerical result. Using now 
relation (\ref{lgxx}), we obtain that the smallest free path length
\begin{equation}
l_{\gamma, {\rm min}} = \frac{2\sqrt{2}}{\sqrt{3}} \, \frac{\cal A}{x_{0}^2} \, R
\label{xi22}
\end{equation}
corresponds to $0.01 \, R \sim R_{0}$.

\begin{table}
\caption{Multiplicity $\lambda$ for $P = 1$ s, $B_{12} = 1$, and $\chi = 30^{\circ}$.}
\begin{tabular}{cccccccccc}
\hline
$x_{0}$ & 0.1 & 0.2  & 0.3 & 0.4 & 0.5 & 0.6 & 0.7 & 0.8 & 0.9 \\
%$\lambda$&  & 0  & 226 & 390 & 512 & 594 & 528 & 228  & 0 \\
$\lambda$ & 0 & 36 & 157 & 305 & 382 & 350 & 209 & 42 & 0 \\
%$\lambda$ &  & 41 & 189 & 379 & 526 & 551 & 412 & 130  & 0 \\
$\lambda_{\rm GR}$ & 7 & 173 & 498 & 829 & 995 & 1040 & 663 & 220 & 3 \\\hline
\end{tabular}
\label{tab3}
\end{table}

Finally, Table~\ref{tab3} shows how the multiplicity $\lambda$ (i.e. the total number of secondary particles $n_{\rm e} = n_{+} + n_{-}$ per one primary particle) depends on the position $r_{0} = x_{0}R_{0}$. As one can see, taking into account the effects of general relativity leads to a significant (by several times) increase in the production rate of secondary particles. Therefore, it seems necessary to include the effects of general relativity in the consideration of the processes of the production of secondary particles near ''the death line''. 

As was already mentioned above, the beginning of the cascade can be initiated by the cosmic gamma background producing $10^5$--$10^8$ primary particles per second in the polar cap region~\citep{SRad82}. If the primary particles have 2D spatial distribution \mbox{${\rm d}N_{\rm prim} = n_{\rm prim}(r_{0},\varphi) r_{0}{\rm d}r_{0}{\rm d}\varphi$,} we obtain for the 2D number density of the secondary pairs $n_{\pm}(r_{\perp}, \varphi)$
\begin{equation}
n_{\pm}  = \frac{\sqrt{3}}{2 \pi} \,\frac{e^2}{\hbar c} \int_{r_{\perp}}^{R_{0}} r_{0}{\rm d}r_{0} \int_{0}^{H}{\rm d}h
\frac{\gamma_{\rm e} F(\omega/\omega_{\rm c})}{R_{\rm c} \omega r_{\perp}} 
    \frac{{\rm d}\omega}{\,\,\,{\rm d}r_{\perp}}n_{\rm prim}.
\label{nescbis}
\end{equation}
Finally we obtain for $n_{\pm}(r_{\perp}, \varphi)$
\begin{equation}
n_{\pm} = \frac{3\sqrt{3}}{16 \pi} \frac{e^2}{\hbar c}  \frac{R_{0}}{R}  \int_{x_{\perp}}^{1}  \frac{x_{0}^2{\rm d}x_{0}}{x_{\perp}(x_{0}-x_{\perp})}
\int_{0}^{H}\frac{{\rm d}h}{R}\gamma_{\rm e}(x_{0},h) F(\xi)
n_{\rm prim}.
\label{nescd}
\end{equation}

\begin{figure}
		\center{\includegraphics[width=0.8\linewidth]{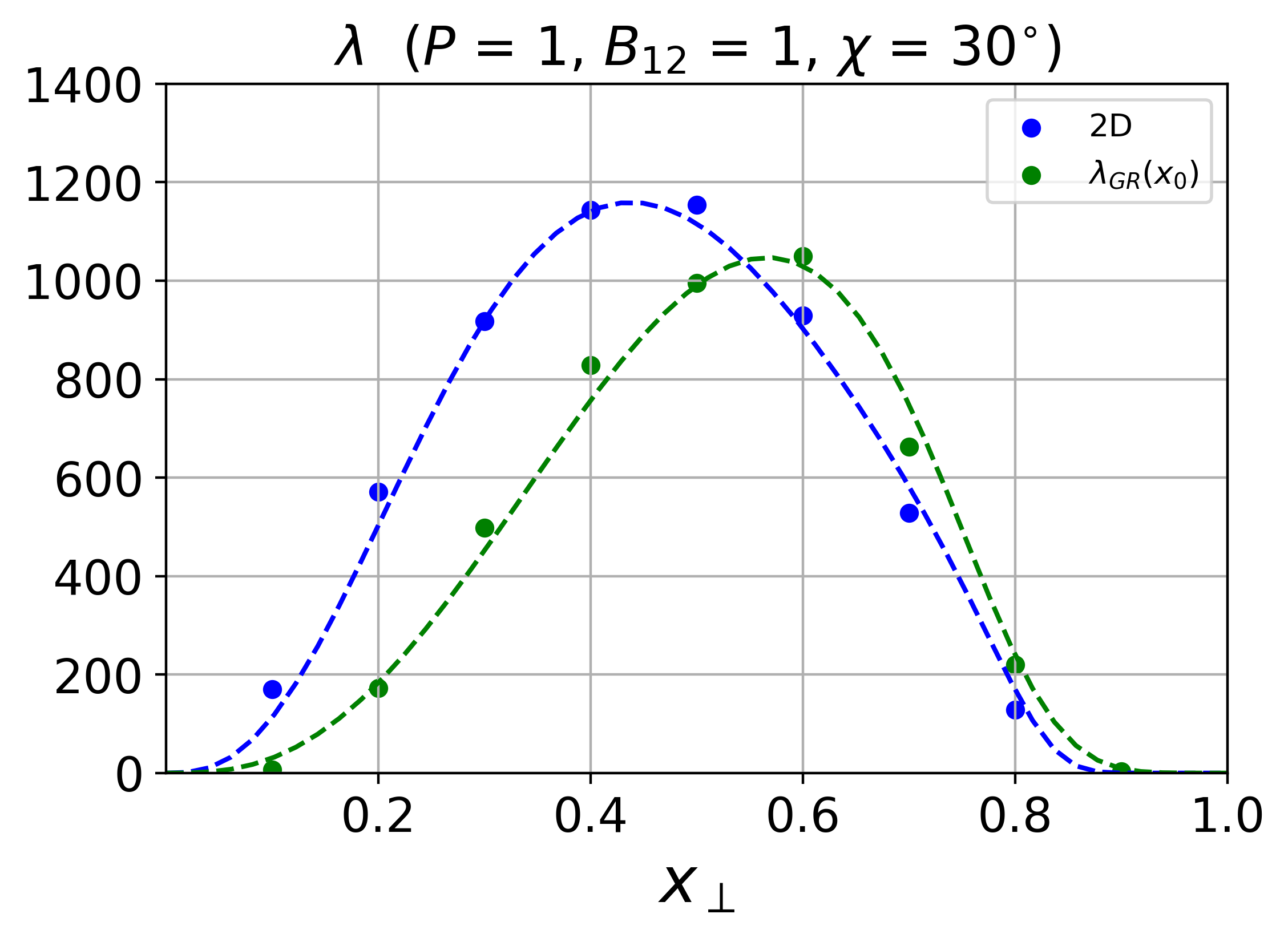}  }
	\caption{2D distribution of secondary particle multiplicity $\lambda(x_{\perp}) = (n_{+}+ n_{-})/n_{\rm prim}$ (\ref{nescd}) generated by the homogeneous primary particle distribution $n_{\rm prim} = 1$ for an ordinary pulsar ($P = 1$ s, $B_{12} = 1$, $\chi = 30^{\circ}$). The right shifted distribution corresponds to multiplicity $\lambda_{\rm GR}(x_{0})$ presented in Table~\ref{tab3}.}
\label{FigA}	
\end{figure}

In Figure~\ref{FigA}, we show the 2D distribution of the effective secondary particle multiplicity $\lambda(x_{\perp}) = (n_{+}+ n_{-})/n_{\rm prim}$ (\ref{nescd}) generated by the primary particles with the homogeneous distribution $n_{\rm prim} = 1$ for an ordinary pulsar ($P = 1$ s, \mbox{$B_{12} = 1$,} $\chi = 30^{\circ}$). The right shifted distribution corresponds to the multiplicity $\lambda_{\rm GR}(x_{0})$ presented in Table~\ref{tab3}. The fitting curves for $x_{\perp} \ll 1$ correspond to asymptotic behaviour $n_{\pm} \propto x_{\perp}^3$ (see Andrianov et al., in preparation), which, as we see, are satisfied with good accuracy. As expected, the 2D distribution $\lambda(x_{\perp})$ is shifted left relative to the distribution $\lambda_{\rm GR}(x_{0})$, since, as was shown in Figure~\ref{FigE1}, secondary particles are born closer to the magnetic axis compared to the position of the primary particle ($x_{\perp} < x_{0}$).

\subsection{Inflow}

Let us note straight away two essential circumstances. At first, for the formation of 
a particle production cascade, secondary electron-positron pairs corresponding 
to the most energetic $\gamma$-quanta must be produced in the region of a 
sufficiently strong longitudinal electric field $E_{\parallel}$ so that one of 
the produced particles can stop and then be accelerated in the opposite direction
(i.e. inwards to the star surface). 
For ordinary pulsars, this occurs at distances from the star's surface
$h \ll R_{\rm cap}$, where the very existence of a longitudinal electric field is 
beyond doubt~\citep{RS, Arons1982}. 

However, near ''the death line'', the free path length $l_{\gamma}$ of 
$\gamma$-quanta propagating outwards become much larger than the size
of the polar cap. Therefore, secondary particles begin to be born at 
the distances from the star surface $h \gg R_{\rm cap}$ where the 
longitudinal electric field, as was previously assumed, practically
vanishes.  However, as was shown, in a real dipole geometry, the 
longitudinal electric field also exists at large distances. It turns 
out that such an additional longitudinal field $E_{\parallel}^{\rm add}$
(\ref{Eadd}) is sufficient to stop the particles at the distances $h \sim R$ 
from the star surface.

Indeed, as one can easily show by passing to a reference frame in which 
$\gamma$-quantum propagates perpendicular to the external magnetic field, 
after an almost instantaneous transition to the lower Landau level, the 
energy of the secondary particles can be written as $\gamma_{\pm}m_{\rm e}c^2$,
where $\gamma_{\pm} = \theta_{\rm b}^{-1}$, i.e.
\begin{equation}
    \gamma_{\pm} \sim \frac{R_c}{l_\gamma}.
\label{gammapm}
\end{equation}
Therefore, before stopping, one of the secondary particles must pass the length
\begin{equation}
    \delta l = \frac{\gamma_{\pm} m_{\rm e}c^2}{eE_{\parallel}}.
\label{deltal}
\end{equation}
It gives
\begin{equation}
    \cfrac{\delta l}{R} \sim \cfrac{R_{\rm c}}{l_\gamma}\cfrac{m_ec^2}{eE_{\parallel}R}
    \approx A^{-1} \cfrac{R}{l_\gamma} \, \cfrac{R^2c^2}{\Omega \omega_{B}R_{\rm cap}^4}
    \left(\cfrac{l}{R}\right)^{1/2},
\label{60}
\end{equation}
where 
$A = 3/16 (f/f_*)(1 - f/f_{*}) \approx 0.1$, $\omega_{B} = eB_{0}/m_{\rm e}c$, 
and we used the relation $R_{\rm c} \approx R/R_{\rm cap}(f/f_{*})^{-1/2}$.
%Because of $l \sim R$,
%$R_0 \simeq \left(\cfrac{\Omega R}{c}\right)^{\frac{1}{2}}R$ and  
%$R_c \simeq \cfrac{R}{\theta} \simeq \left(\cfrac{\Omega R}{c}\right)^{\frac{1}{2}}R $:
As a result, we obtain
\begin{equation}
    \cfrac{\delta l}{R} \sim  A^{-1} \frac{R}{l_{\gamma}} \,
    \cfrac{c}{\omega_B R}\left(\cfrac{\Omega R}{c}\right)^{-3} \sim 10^{-2} P^{3}B_{12}^{-1},
\label{dlR}
\end{equation}
so that the stop length $\delta l$ is, indeed, much less than the free pass 
$l_{\gamma} \sim R$. Therefore, in what follows, we assume that one of 
the secondary particles begins its reverse motion at the point of its birth.

Here, however, one should pay attention to the fact that the additional 
longitudinal electric field $E_{\parallel}^{\rm add}$ works effectively 
only on that part of the polar cap which is located closer to the rotation 
axis ($\sin\varphi > 0$ for $\chi < 90^{\circ}$ and $\sin\varphi < 0$ 
for $\chi > 90^{\circ}$). In the other parts of the open field lines, 
where the opposite inequalities are made, the direction of the additional 
longitudinal electric field will be opposite to the electric field near the 
star surface. In particular, for $\sin\varphi = 0$, there is no additional 
longitudinal electric field $E_{\parallel}^{\rm add}$ at all.

Finally, we note one more circumstance which significantly distinguishes the 
processes of pair creation by $\gamma$-quanta moving towards the star compared 
to the case of moving from the star's surface considered above. The point is 
that, as shown in Figure~\ref{FigD}, the particles moving 
from the surface of a star are accelerated on a scale $R_{0} \ll R$, so that 
the free path length of $\gamma$-quanta can be comparable to the radius of 
the star. On the other hand, the particles moving towards the star gain most
of the energy only near the very surface. Therefore, the free path length of 
$\gamma$-quanta should be of the order of $R_{0}$.

As was shown above, the energy of the particles ${\cal E}(h)$ 
moving towards the star can be written with good accuracy in the form
\begin{equation}
{\cal E}(h) = e \, \delta \psi(h),
\label{b1}
\end{equation}
where, according to (\ref{psipsi}),
\begin{equation}
\delta \psi(h) = \frac{1}{2} \, \frac{\Omega B_{0}R_{0}^2}{c} {\cal P}(r_{m}, \varphi_{m})f(h). 
\label{b2}
\end{equation}
Here,
\begin{equation}
{\cal P}(r_{m}, \varphi_{m}) = \left(\cos\chi + \frac{3}{4}\frac{r_{m}}{R}\sin\chi\cos\varphi_{m}\right) \left(1 - \frac{r_{m}^2}{R_{0}^2}\right),
\label{b3}
\end{equation}
and
\begin{equation}
f(h) \approx \exp\left(- \lambda_{1} h/R_{0}\right),
\label{b4}
\end{equation}
where $\lambda_{1} \approx 2.5$.
The ability to replace the sum of the power functions $(r/R)^{-\lambda/\theta_{0}}$
in (\ref{7bis})
with the exponential term $\exp(-\lambda_{1} h/R_{0})$ is shown in 
Figure~\ref{FigC}.

Using the expression for the free pass length $l_{\gamma}$ (\ref{47}) with the
energy of the curvature $\gamma$-quanta 
\begin{equation}
\hbar \omega_{\rm c} = \frac{3}{2} \xi \, \frac{\hbar c}{R_{\rm c}} \left(\frac{{\cal E}(h)}{m_{\rm e}c^2}\right)^3
\label{b5}
\end{equation}
and the correction factor $\xi$ from (\ref{45''}), we can finally write down the condition 
of the pair creation over the star surface
\begin{equation}
l_{\gamma}(h) < h,
\label{b6}
\end{equation}
which can be rewritten as
\begin{equation}
x > {\cal B} \exp(3 \lambda_{1} x).
\label{b7}
\end{equation}
Here $x = h/R_{0}$,
\begin{equation}
{\cal B} = \frac{2048}{81 \Lambda \xi f_{\ast}^{9/2}} \frac{R^2}{\lambdabar^2}\frac{c^4}{\omega_{B}^4 R^4}\frac{1}{x_{0}^2} \left(\frac{\Omega R}{c}\right)^{-15/2}{\cal P}^{-3},
\label{b8}
\end{equation}
and we again use standard definition (\ref{fstar}) for the dimensionless polar cap area $f_{\ast}$. 
If inequality (\ref{b6}) is violated, then the free path length $l_{\gamma}$ becomes greater than the height above the surface of the neutron star $h$ at which it was generated. In this case, $\gamma$-quantum does not have time to give birth to a pair before it collides with the surface.

As a result, the condition in which secondary particles will be produced both for the primary particles accelerated from the stellar surface and for the opposite motion) can be written down as{{\footnote{When ${\cal B} = 0.05$, we have a single root at $h = 0.13\, R_{0}$.}} 
\begin{equation}
{\cal B} < 0.05,
\label{b9}
\end{equation}
which gives the following evaluation for the maximum period $P_{\rm max}$} 
\begin{equation}
P_{\rm max} = 0.8 \, B_{12}^{8/15}x_{0}^{4/15} {\cal P}^{2/5} \, s.
\label{b10}
\end{equation}
A value of 0.8 seconds corresponds to $R = 12$ km, \mbox{$\Lambda = 15$,} $\xi = 3$ and $f_{\ast} = 1.6$. As one can see, a rather strong dependence of the limiting period $P_{\rm max}$ on the magnetic field $B_{12}$ makes it easy to explain the observed periods in the range of several seconds. A detailed study of this issue, as was already noted, will be carried out in Paper II.

\section{Discussion}
\label{Sect5}

Thus, in this Paper I, the first step was taken in a consistent analysis of the conditions leading to the cessation of the secondary pair production for a sufficiently large rotation period $P$ (which, in turn, leads to the termination of radio emission). As is well-known, in reality, we have not ''the death line'', but ''the death valley'' on the $P-{\dot P}$ diagram, which, apparently, is related to the tail of the distribution on some physical parameters. What parameters determine this band will be the subject of Paper II.

As for Paper I, we started from our main assumption that near ''the death line'', the polar regions are almost completely free of plasma. This made it possible to accurately determine the potential drop in the area of open field lines. In particular, it was shown that in a dipole magnetic field in the region of open field lines, there is a longitudinal electric field which can stop secondary particles even at sufficiently large distances $h \sim R$ from the surface of the star. To date, this property has not been known.

In addition, the corrections related to the effects of general relativity were determined. As expected, they turned out to be at the level of 10--20 per cent, i.e. at the same level of uncertainty which can be assumed in other quantities, such as the radius and  moment of inertia of a neutron star. Nevertheless, due to the strong dependence of the production rate on the energy of primary particles, taking into account the effects of general relativity leads to a significant (several times) increase in the multiplicity of particle production $\lambda$.  Therefore, in this work, they were included into consideration. Looking into the future, we immediately note that it can be expected that the spread in such quantities as the radius of the star $R$ and the size of the polar cap  $R_{0}$ (and, certainly, the curvature radius of magnetic field lines $R_{\rm c}$) should also lead to a noticeable expansion of ''the death line''. Paper II will be devoted to the analysis of all these issues.

Further, the question of the spatial distribution of the secondary particles produced by curvature photons was investigated in detail (as to synchrotron photons, we assume that near ''the death line'', they are not efficient in the production of secondary pairs). First of all, it was shown that a certain role in the production of secondary plasma can play $\gamma$-quanta, the energy of which is several times higher than the energy of the maximum of the spectrum $0.44 (\hbar c/R_{\rm c})\gamma^3$. This effect becomes especially important for the curvature $\gamma$-quanta propagating towards the neutron star surface. As a result, conditions (\ref{b10}) for the termination of the cascade were formulated quantitatively. A detailed analysis of the dependence of ''the death line'' both on the parameters of a neutron star and on the possible existence of a nondipole magnetic field is to be carried out in Paper II.

\section*{Data availability}
The data underlying this work will be shared on reasonable   request to the corresponding author.

\section*{Acknowledgements}
Authors thank Ya.N.Istomin and A.A.Philippov for useful discussions. This work was partially 
supported by Russian Foundation for Basic Research (RFBR), grant 
20-02-00469. 

\bibliographystyle{mnras}
\bibliography{Beskin-PaperI}

\appendix

\section{Approximation by Bessel functions}
\label{A1}

\begin{figure}
	\begin{minipage}{0.9\linewidth}
		\center{\includegraphics[width=1.0\linewidth]{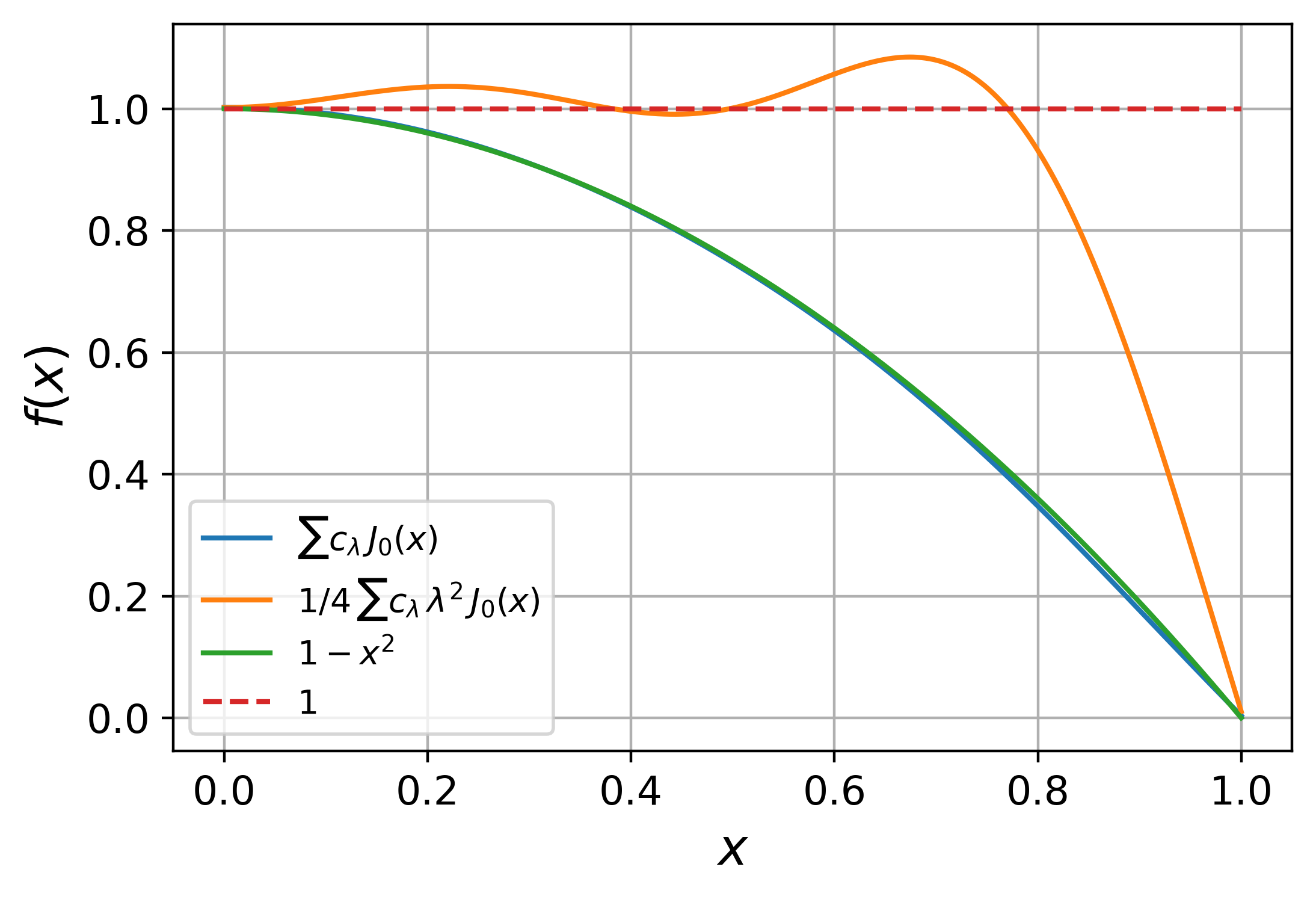}  }
	\end{minipage}
	\hfill
	\begin{minipage}{0.9\linewidth}
		\center{\includegraphics[width=1.0\linewidth]{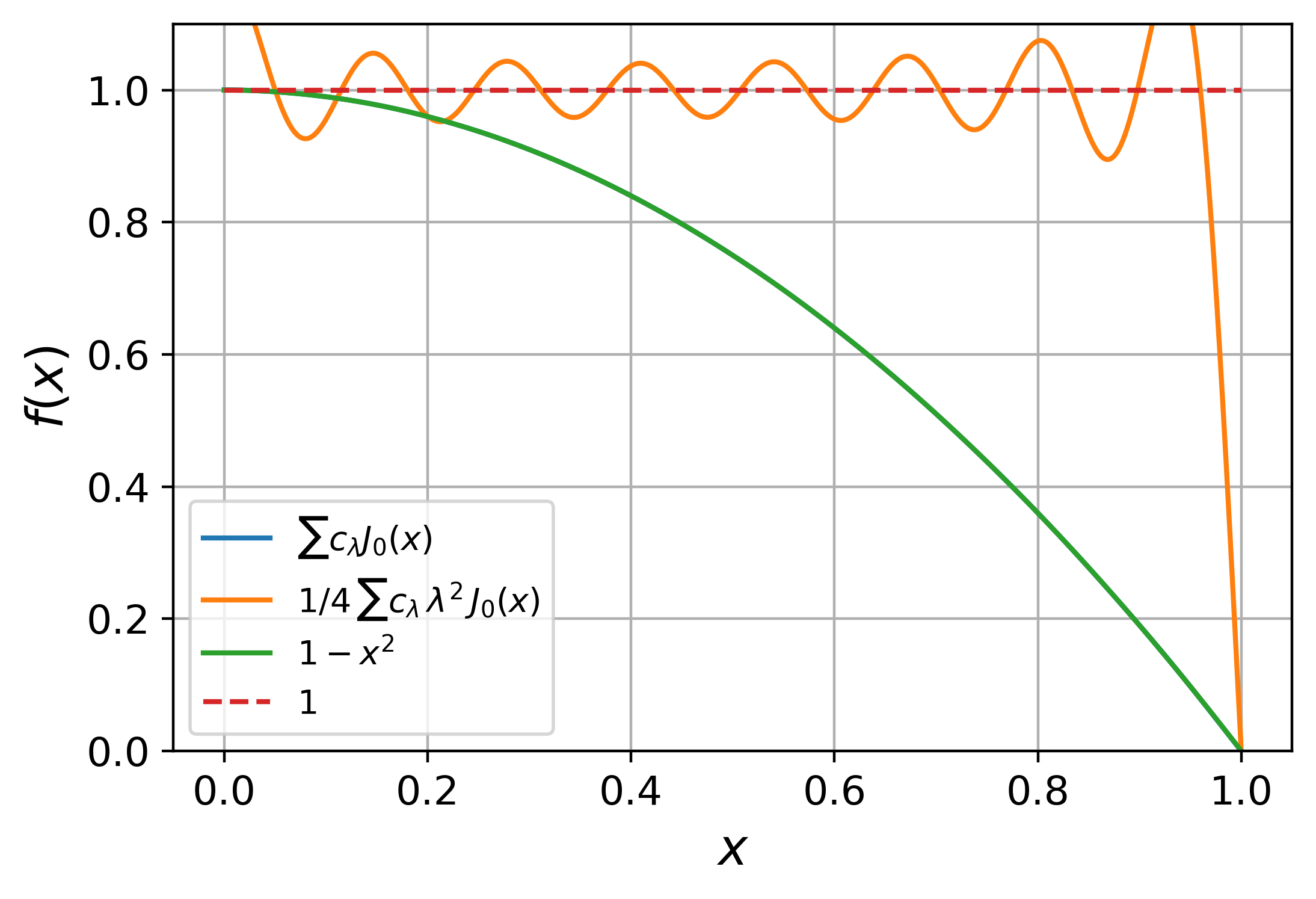}  }
	\end{minipage}
	\caption{Expansion of the function $1 - x^2$ by Bessel functions 
	$J_{0}(\lambda_{i}^{(0)}x)$ (\ref{sum1}) using our four terms 
	$c_{i}^{(0)}$ (\ref{ccc}) (top) and the first exact twenty 
	terms (\ref{ccc1}) (bottom). The degree of agreement of the 
	second derivative (\ref{c01}) is also shown.}
\label{FigA0}	
\end{figure}

\begin{figure}
	\begin{minipage}{0.9\linewidth}
		\center{\includegraphics[width=1.0\linewidth]{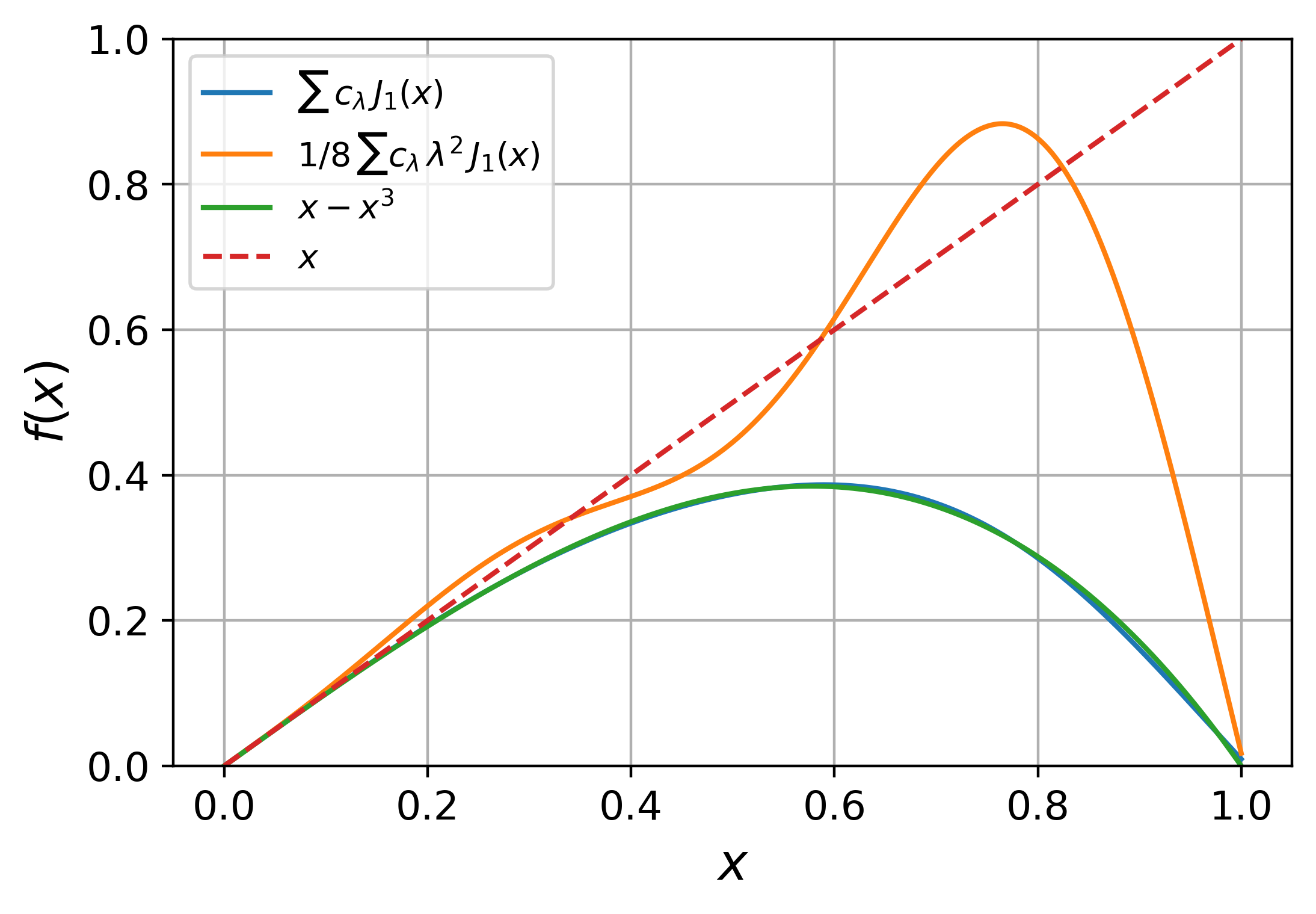}  }
	\end{minipage}
	\hfill
	\begin{minipage}{0.9\linewidth}
		\center{\includegraphics[width=1.0\linewidth]{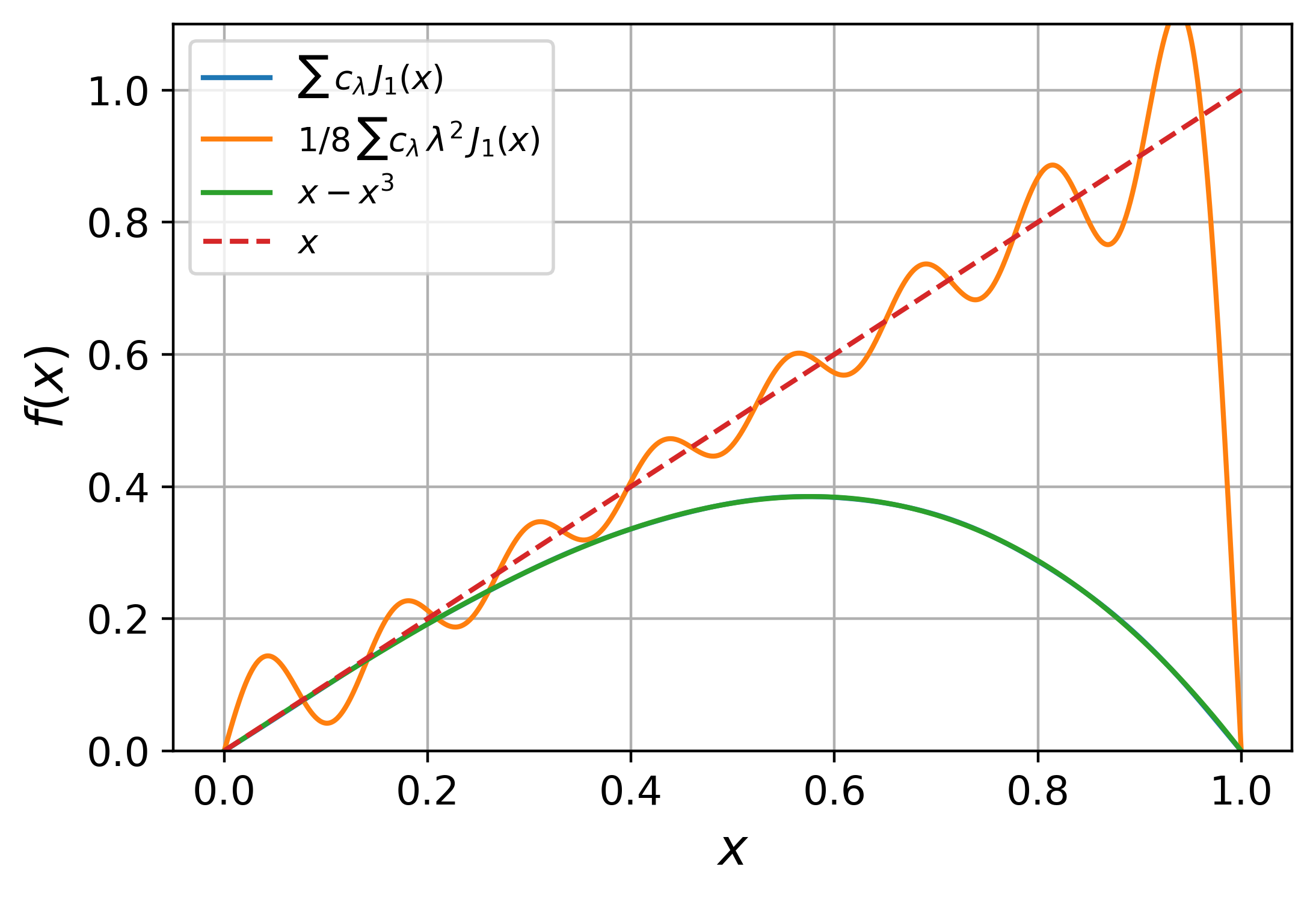}  }
	\end{minipage}
	\caption{Expansion of the function $x - x^3$ by Bessel functions 
	$J_{1}(\lambda_{i}^{(1)}x)$ (\ref{sum2}) using our four terms 
	$c_{i}^{(1)}$ (\ref{cc}) (top) and the first exact twenty terms 
	(\ref{ccc2}) (bottom). The degree of agreement of the second derivative 
	(\ref{c11}) is also shown.}
\label{FigA1}	
\end{figure}

In this Appendix, we illustrate various aspects related to series 
expansion in the Bessel functions discussed in section~\ref{Sect1}.
In Figures~\ref{FigA0}--\ref{FigA1} we show how the expansions in Bessel 
functions (\ref{sum1}) and (\ref{sum2}) approximate the functions 
$1 - x^2$ and $x - x^3$. The top pannels show an approximation when we 
restrict ourselves to only the first four terms $c_{i}$ (\ref{ccc}) and 
(\ref{cc}), for which the coefficients $c_{1}$  and $c_{2}$ coincide with 
the precisely calculated values
\begin{eqnarray}
c_{i}^{(0)} & = & \frac{2}{J_{1}^{2}(\lambda_{i}^{(0)})} \,
\int_{0}^{1} x(1-x^2)J_{0}(\lambda_{i}^{(0)}x){\rm d}x, 
\label{ccc1}\\
c_{i}^{(1)} & = & \frac{2}{J_{2}^{2}(\lambda_{i}^{(1)})} \,
\int_{0}^{1} x(x-x^3)J_{1}(\lambda_{i}^{(1)}x){\rm d}x,
\label{ccc2}
\end{eqnarray}
and the two remaining ones are selected to satisfy relations
(\ref{sum1}) and (\ref{sum2}). Accordingly, the approximations 
by the first twenty terms of the series are shown in the bottom ones.
As we see, both series reproduce the functions $1 - x^2$ and 
$x - x^3$ with high accuracy. As for their second derivatives
(\ref{c01}) and (\ref{c11}), in some respects, our first four 
terms approximate these functions even better than twenty 
exact terms.

\begin{figure}
		\center{\includegraphics[width=0.95\linewidth]{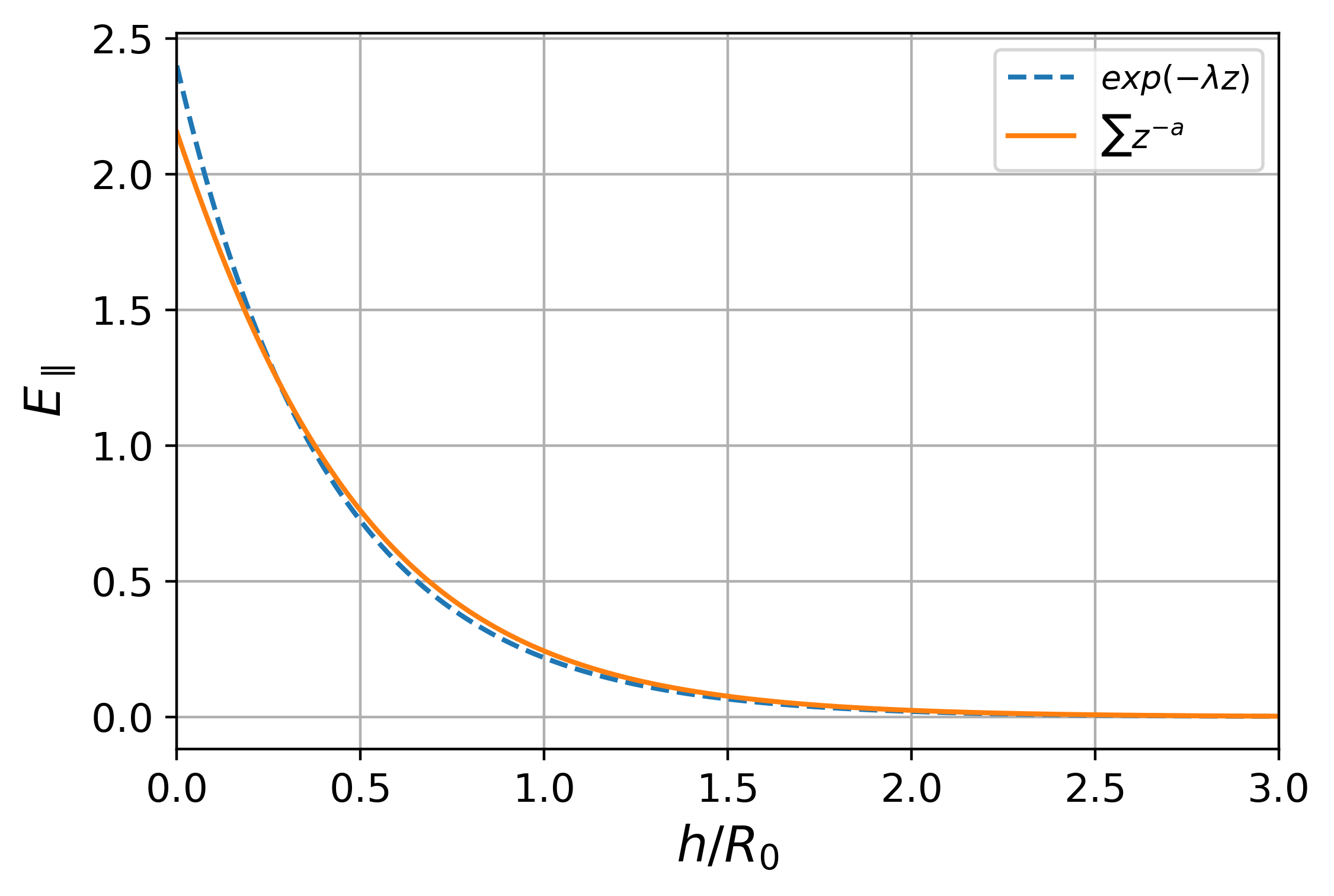}  }
	\caption{Dimensionless parallel electric field $E_{\parallel}$
	as a function of $z = h/R_{0}$ determined by a power series 
	(\ref{7bis}) (solid line) and by one single exponential term 
	(\ref{b4}) (dashed line).}
\label{FigC}	
\end{figure}

Finally, in Figure~\ref{FigC}, we present the comparison of the dimensionless parallel component 
of the electric field \mbox{$E_{\parallel} = -(\nabla \psi {\bf B})/B/(\psi_{0}/R_{0})$} 
calculated for cylindrical and conical geometries for $P = 1$ s, $B_{12} = 1$, and 
$\chi = 0^{\circ}$. As we can see, although in the first case, 
the expansion is carried out in exponential functions $e^{-\lambda_{i}z}$, and for 
dipole geometry in power-law dependencies $(r/R)^{-\lambda_{i}/\theta_{0}}$, their 
difference becomes indistinguishable already for four first terms of the series. 
Moreover, good enough agreement is achieved using only the first
 exponential term 
(dash line). In turn,  this confirms the possibility to write down the 
condition for the presence of a cascade of secondary plasma production (i.e. the 
condition that secondary particles will be produced both for primary particles 
accelerated from the stellar surface and for the opposite motion) as (107).

\end{document}